\documentclass{aastex}
\usepackage{emulateapj5}

\def\deg{\hbox{$^\circ$}}
\def\etal{\it et al.}

\slugcomment{Submitted to {\it The Astrophysical Journal}}
\begin{document}

\title{The QMAP and MAT/TOCO  Experiments for Measuring
 Anisotropy in the Cosmic Microwave Background}

\author{A. Miller\altaffilmark{1,2,3},
J. Beach\altaffilmark{1,13},
S. Bradley\altaffilmark{1},  
M. J. Devlin\altaffilmark{4,14},
R. Caldwell\altaffilmark{1,4,12},
H. Chapman\altaffilmark{4},
W.B. Dorwart\altaffilmark{1}, 
T. Herbig\altaffilmark{1,2,5},
D. Jones\altaffilmark{4},
G. Monnelly\altaffilmark{1,6}, 
C. B. Netterfield\altaffilmark{1,7},
M. Nolta\altaffilmark{1}, 
L. A. Page\altaffilmark{1}, 
J. Puchalla\altaffilmark{4,8},
T. Robertson\altaffilmark{1,10}, 
E. Torbet\altaffilmark{1,9},
H. T. Tran\altaffilmark{1}
W. E. Vinje\altaffilmark{1,11}
}

\altaffiltext{1}{Dept. of Physics, Princeton University, Jadwin Hall,
Princeton, NJ 08544} 
\altaffiltext{2}{Hubble Fellow}
\altaffiltext{3}{Dept. of Astronomy and Astrophysics, The University 
of Chicago, Chicago, IL 60637}
\altaffiltext{4}{Dept. of Physics and Astronomy, The University of 
Pennsylvania, Philadelphia, PA 19104}
\altaffiltext{5}{McKinsey \& Co., 3 Landmark Square, Stamford, CT 06901}
\altaffiltext{6}{MIT CSR, Bldg 37-524, 70 Vassar St. Cambridge MA, 02139}
\altaffiltext{7}{Dept. of Astronomy, University of Toronto, Toronto, 
Ontario, M5S 1A7} 
\altaffiltext{8}{Dept. of Molecular Biology, Princeton University, 
Princeton, NJ 08544}
\altaffiltext{9}{Dept. of Physics, University of 
California, Santa Barbara, CA 93106}
\altaffiltext{10}{Dept. of Physics, University of California at
Berkeley, Berkeley, CA 94720}
\altaffiltext{11}{Neurobiology Program, University of California at
Berkeley, Berkeley, CA 94720}
\altaffiltext{12}{Dept. of Physics, 6127 Wilder Lab, 
Dartmouth College, Dartmouth, NH 03755}
\altaffiltext{13}{Xerox Corporation, Palo Alto, CA  94304}
\altaffiltext{14}{Sloan Fellow}

%
%
%
%
%

\begin{abstract}
We describe two related experiments that measured the anisotropy in the 
cosmic microwave background (CMB). 
QMAP was a balloon-borne telescope that flew twice in 1996,
collecting data on degree angular scales with an array of 
six high electron mobility transistor-based amplifiers (HEMTs).
QMAP was the first experiment to use an interlocking scan strategy 
to directly produce high signal-to-noise CMB maps.
The QMAP gondola was then refit for ground based work as the
MAT/TOCO experiment. Observations were made from 5200 m on
Cerro Toco in Northern Chile in 1997 and 1998 using time-domain
beam synthesis. MAT/TOCO was the first experiment to see both 
the rise and fall of the CMB angular spectrum, thereby localizing the 
position of the first peak to $l_{peak}=216\pm14$. In addition to describing
the instruments, 
we discuss the data selection methods, checks for systematic
errors, and we compare the MAT/TOCO results to those from recent
experiments. We also correct the data to account for 
an updated calibration and 
a small contribution from foreground emission. We find the 
amplitude of the first peak for $160<l<240$ to be 
$\delta T_{peak}=80.9\pm3.4\pm5.1~\mu$K,
where the first error is statistical and the second is from calibration.

\end{abstract}

\section{Introduction}
Experiments aimed at measuring the anisotropy in the CMB require
a combination of sensitive detectors and novel observing strategies. The
observational goal is to measure micro-Kelvin celestial 
variations in thermal emission with a telescope observing from an 
environment that is some ten million times brighter. 
Below 90 GHz, the detectors of choice have been 
high electron mobility transistor based amplifiers designed at NRAO
[HEMTs, Pospieszalski (1992)]. Above 90 GHz, bolometers are the best detectors 
[e.g., \cite{bock98,Lee96,downey,Tucker91}]. SIS-based systems
\cite{Kerr93} near 100 GHz have the
speed and intrinsic sensitivity of transistor amplifiers though do 
not yet have the
large instantaneous bandwidth of bolometers or HEMTs. 
Over the past five years, instruments have been designed for direct 
mapping of the CMB 
[e.g.,  QMAP, BOOMERanG \cite{boominst}, MAXIMA \cite{Hanany00}, 
TopHat (2001) ], and for beam synthesis
[Saskatoon (SK) \cite{Wollack97}, MSAM \cite{Fixsen96},   Tenerife/Bolo
\cite{Romeo00}, PYTHON \cite{Coble}, VIPER \cite{Peterson00}]. 
More recently, interferometers
based on HEMT amplifiers
have reported CMB anisotropy results [CAT \cite{bak99}, DASI \cite{Leitch01}, 
IAC \cite{Harrison00}, CBI \cite{pad00}].
Though the primary data product of the interferometers and beam-synthesis
experiments is the angular spectrum, data taken with these techniques
can be turned into maps (e.g., Tegmark 1997). Conversely, there is always 
some filtering 
involved in the mapping experiments. A common element of these 
experiments is that they are limited by systematic error. 

In this paper, we describe the instruments for the QMAP experiment and
the Mobile Anisotropy Telescope on Cerro Toco (MAT/TOCO or TOCO for
short). We supply the details necessary for assessing the quality of the
data and reproducing the experimental method.
QMAP is described in part in Devlin {\it et al.} (1998) and analyses 
of the data are presented in Herbig {\it et al.} (1998), de 
Oliveira-Costa {\it et al.} (1998, 1999), Xu {\it et al.} (2000), 
and Park {\it et al.} (2001). Balloon borne
mapping experiments have a long history \cite{weiss,partridge} though
highly interlocking scan strategies over limited regions of sky
are more recent \cite{Staren99,dB00,Lee01}. QMAP, which flew twice in
1996, was the first of these to produce a ``true map'' of the CMB,
complete with pre-whitening and full covariance matrices. 
QMAP was comprised of a focal plane array
of three dual polarized HEMT channels with an angular resolution of
roughly $0\fdg8$. The beam array was steered on the sky
by a large chopping flat. 

\begin{figure*}[tb]
\epsscale{0.75}
\plotone{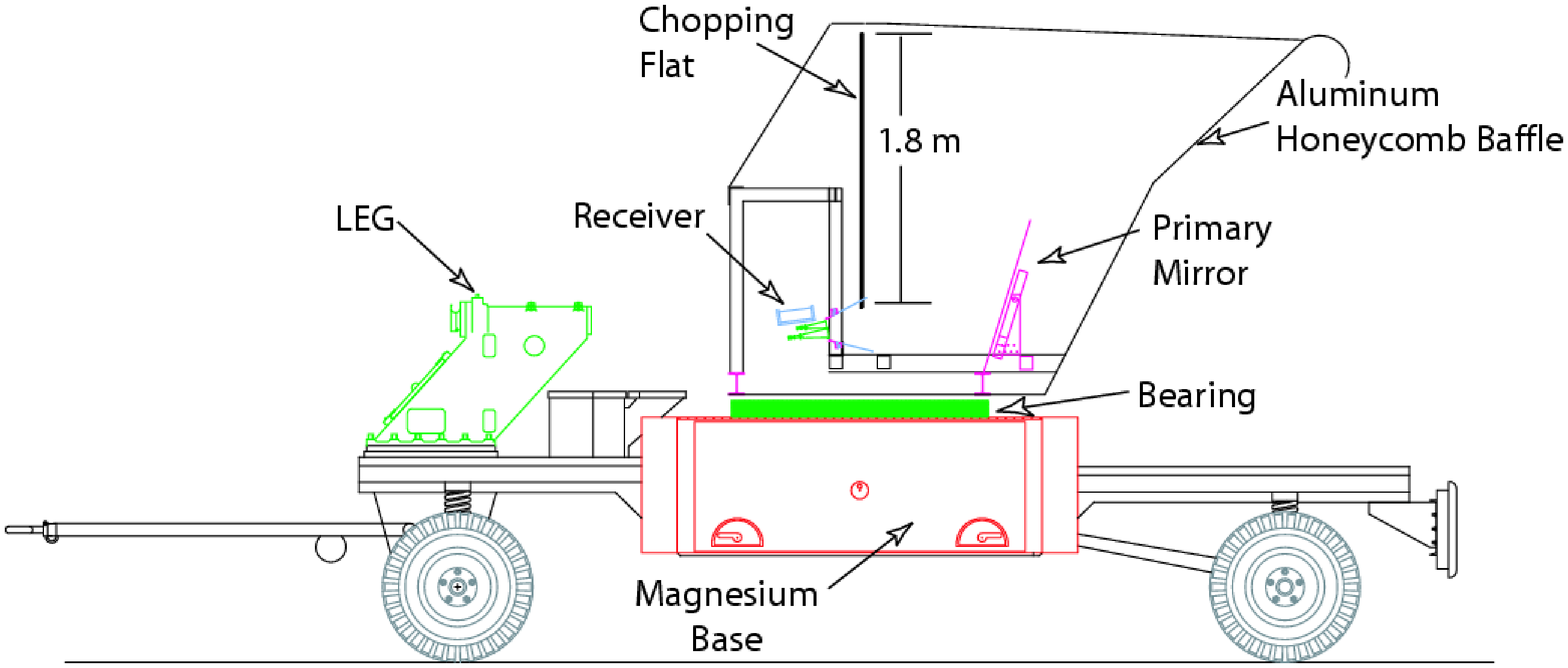}
\caption{Schematic of the Mobile Anisotropy Telescope. The legs are
shown in their stowed position on the left end of the trailer. When 
observing, the compressor is also on the trailer. The QMAP gondola 
comprises everything above the bearing.}
\label{fig:1mat_schematic}
\end{figure*}

TOCO used the QMAP gondola and receiver refit to work with a 
mechanical cooler instead of liquid cryogens. It also employed two
SIS-based\footnote{SIS stands
for Superconductor-Insulator-Superconductor. The detecting element is a
quasi-particle mixer \cite{Tucker85}.} 144 GHz detector systems 
to improve the resolution
to $0\fdg2$. TOCO employed the Saskatoon-style
beam synthesis strategy \cite{net95} with eight independent detectors.
Instead of observing near the NCP from Saskatoon, Canada, we observed near 
the SCP from the side of Cerro Toco in Northern Chile
\footnote{ The Cerro Toco site of
the Universidad Cat\'olica de Chile was made available through the
generosity of Prof. Hern\'an Quintana, Dept. of Astronomy and
Astrophysics. It is near the ALMA site.}.
At 144 GHz, the atmospheric column density in Saskatoon is too large 
for anisotropy measurements;
a high altitude site such as the Chilean Altiplano is required.
TOCO operated for two seasons in 1997 and 1998. The primary results
and short description of the instrument are given in 
Torbet {\it et al.} (1999) and
Miller {\it et al.} (1999). The $0\fdg2$ resolution allowed us to 
locate the first peak in the angular spectrum at $l\approx 212\pm14$
(Knox and Page 2000). In the context of the popular adiabatic CDM models,
this shows that the universe is geometrically flat 
\cite{doris78,kam94,Bond94,HW96,Cornish00}. 

\section{Overview of Gondola and Mobile Anisotropy Telescope}

The TOCO experiment is shown schematically in
Figure~\ref{fig:1mat_schematic}\footnote {The Nike Ajax radar 
trailer was donated to the
University of Pennsylvania by Lucent Technologies.}. 
The part of the figure containing the optics and receiver is the 
QMAP balloon gondola. The
radar trailer, on which the gondola is mounted, has a
separable magnesium base. Three legs hold the
base off the trailer and stabilize it. When the telescope
is transported, the legs are removed. The radar trailer
has a 1.4 m diameter precision bearing 
on which is mounted a 2.5 cm thick
flat plate that holds the gondola. The
plate and gondola are rotated using an on-axis DC motor\footnote
{The motor is a Compumotor DR 1100A - 100~Nm torque.}.
CMB observations are made with the telescope in a
fixed position; a brake holds the telescope in place allowing the 
motor to be shut off during observations and preventing it from 
drawing large currents or oscillating as it seeks the target position
in high winds. The motor has a 15 cm diameter hole
in the center through which
cables and refrigerator hoses pass from the inside of the
telescope to the outside. The compressor that runs the mechanical 
cryocooler is mounted
on the trailer. The azimuth is instrumented with an absolute
17-bit encoder and a 20-bit resolver.

\section{The Receiver}

\begin{figure*}[tb]
\epsscale{0.5}
\plotone{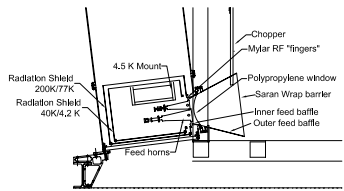}
\caption{View of the inside of the dewar. Shown are the 
4.5~K mounting structure, the thermal shield and cold plate at 40~K, and the 
shield at $\approx 200$~K. In the QMAP configuration, the entire cold plate
was cooled in the lab to 4.2~K with liquid Helium, or to
$\approx 2.7$ K in flight ($\approx
33,000$ m altitude). Pressure regulated LN$_2$ held the intermediate
temperature stage near 77~K. The SIS mixers are mounted to the 4.5~K
structure. There is a 
small section removed from the bottom of the chopper to 
accommodate the outer feed baffle. One source of modulated radiation 
is the cavity formed by the moving chopper and the feed baffle. }
\label{fig:dewar_shells}
\end{figure*}

Radiation from the sky enters the dewar through a 15.25 cm diameter vacuum
window made of 0.56 mm polypropylene and is collected with corrugated 
feed horns as shown in Figure~\ref{fig:dewar_shells}. Three aluminum
baffles define the entrance aperture, one is attached to the 40 K cold plate,
one is attached to the dewar just inside the vacuum window, and one is attached
to outside of the dewar. Strips of
aluminized Mylar connecting the cold feeds to the ambient temperature
dewar block RF interference and reduce
optical loading on the cold stage. To prevent the formation of frost on
the window, warm air is passed through a volume in front of the vacuum
window defined by a Saran Wrap-covered aluminum cone attached to the 
outside of the dewar.

\begin{figure*}[tb]
\epsscale{0.9}
\plotone{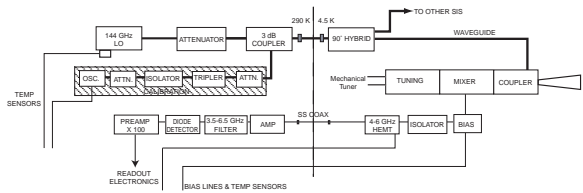}
\caption{Components of the SIS Receiver. The components on the right are
in the dewar. Components on the left are housed in the thermally
stabilized 293~K backpack. The temperatures of all critical components are
monitored.}
\label{fig:siscon}
\end{figure*}

Two NRAO SIS mixers are attached to the $\approx 4.5~$K stage of the
dewar and six HEMTs are attached to the 40~K stage,
two with center frequencies of 31 GHz (in Ka band) and four with
center frequencies of 42 GHz (in Q band). Warm amplifiers, bandpass filters, noise
sources, and local oscillator for the SIS system are housed in a 293~K
``backpack'' attached to the outside of the dewar. The primary
difference between the TOCO and QMAP receiver configurations is that
QMAP used liquid cryogens and TOCO used a mechanical refrigerator to
cool the HEMTs as well as the SIS mixers.

\begin{figure*}
\epsscale{0.8}
\plotone{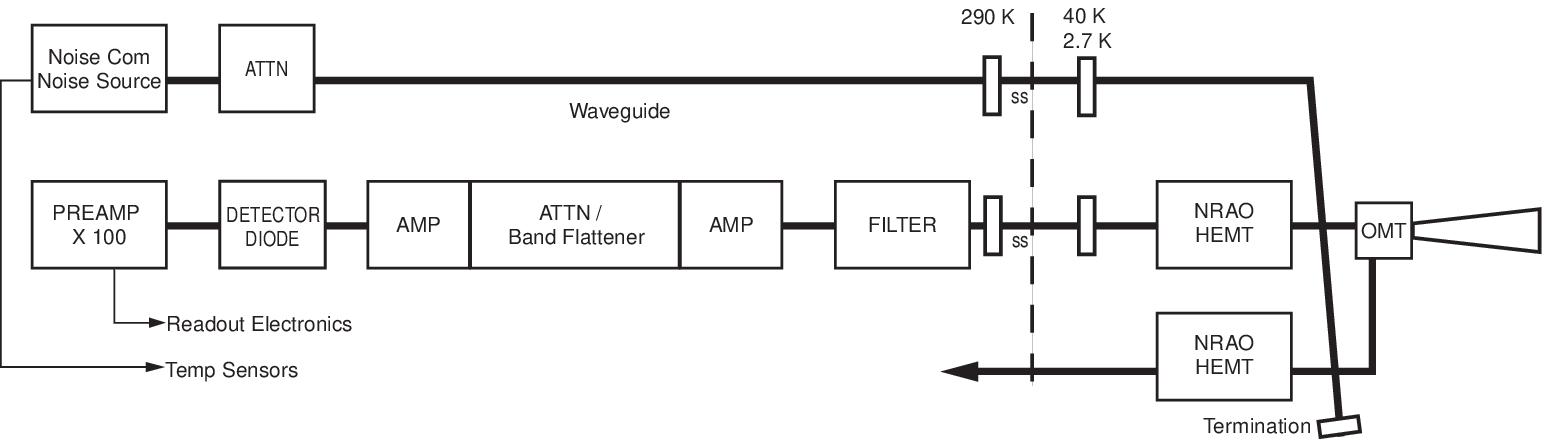}
\caption{\small Components of a Q-band receiver, one of six HEMT radiometer channels.
Components on the right are mounted to the 40 K (for TOCO, 2.7~K for
QMAP) plate inside the dewar, and components on the left are 
in the 293~K backpack. The HEMT output is carried on a
rectangular copper waveguide to the edge of the cold stage where it is
connected to 0.5~m long thin walled stainless steel waveguide that 
runs down the vacuum chamber inside the dewar to warm stage. 
The stainless steel waveguide provides thermal isolation between the
cold stage and ambient temperature. Copper waveguide from 
the ambient temperature end of the 
stainless steel segment runs to 
the vacuum seal which is formed with a piece of 0.013 mm thick kapton
tape.}
\label{fig:hemt}
\end{figure*}

\subsection {The SIS Configuration}

The SIS configuration is shown in Figure~\ref{fig:siscon}.
The mixers are coupled to the sky with conical corrugated feed horns.
A round-to-rectangular
transition at the base of the horn transforms the incoming
signal to a single polarization which is added to a
signal from the local oscillator using a 20 dB branch line
coupler. The combined signals are fed into the SIS mixer block. 
A sliding backshort,
connected to the outside of the dewar
by a flexible shaft, is used to tune the SIS mixer, while cold, for
optimal coupling. The SIS is biased through a bias T 
which allows transmission of the RF signal while blocking
the DC bias voltage. An
associated circuit controls and reads the current through
and voltage across the SIS. The signal then comes out of the
bias T, goes through a 20 dB 4-6~GHz isolator (P \& H Labs) and into a cold 
C-band (3.95-5.85 GHz) HEMT amplifier with 33~dB gain, as diagrammed in
Figure 3. The SIS,
branch line coupler, 
bias T, and C-band HEMT amplifier were all designed and fabricated at
the National Radio Astronomy Observatory (NRAO). 

The output of the cold HEMT is carried on stainless steel 
semi-rigid coax from the
cold stage to the backpack where it is amplified by a warm 44~dB RF
amplifier, filtered through 
a 3.5-6.5 GHz filter and
detected with a detector diode (Hewlett-Packard: $\pm 0.2$ dB flatness
and a typical sensitivity of 300 mV/mW).
To increase the linearity and to bring the detector output into a
convenient range ($\approx 0.01~$V), the diode is shunted with a 
$\approx1$~k$\Omega$ resistor. The RF filter ensures that only RF
radiation in the passband of the C-band IF amplifier 
makes it through the system. The detected output is then
amplified by 100 and sent to the data acquisition system. 
The IF system
temperature was measured in the lab with a heatable $50~\Omega$ load and
found to be $T_{\rm sys}({\rm IF})=<6$K. The net gain of the system is
73~dB. 
 
The LO (Millimeter Wave Oscillator Company) consists of a
cavity stabilized oscillator with an
output frequency of 48 GHz which injection-locks a second oscillator,
which in turn injection locks a high power Gunn oscillator. The
oscillator feeds a tripler with an output frequency
of 144 GHz. This signal then goes through an attenuator, into a 3 dB
directional coupler and is carried by WR-6 waveguide
to the cold stage. On the cold stage a $90^{\circ}$ hybrid splits the
LO power and sends it to the two mixers. 
The third port of the coupler is used to couple test and calibration
signals into the system. The
total LO power is about 8.5 mW,  $\approx300~$nW of which impinges
on each SIS mixer. 

\subsection {The HEMT Configuration} 
 
Figure~\ref{fig:hemt} shows the configuration of the HEMT receivers
for the TOCO 1997 and 1998 observing
seasons. The QMAP 
configuration is almost identical. 
The radiation enters the dewar through the vacuum aperture
described above and is collected using Ka- and Q-band corrugated
horns. An orthomode transducer (OMT) splits the radiation into vertical and
horizontal linear polarizations. The Ka-band OMT has a reflection
coefficient of less than $-26~$dB over the bandpass of the 
channel. The
Q-band OMTs have a reflection coefficient of less than $-20~$dB. 
Each polarization is  carried by rectangular
copper waveguide to a separate low noise NRAO
cryogenic HEMT amplifier\cite{posp92,posp94}. 
The passband of the system is set by the amplifiers and waveguide cutoff
at the low end and a filter at the high end.
Bends in the waveguide
are limited to a radius of curvature of at least 1.5~cm in order to
minimize reflections \cite{Monnelly96}. 

At the input to each HEMT amplifier is a 20 dB crossed-guide Bethe hole
coupler that is used to inject a noise pulse. There is one source for
all four of the Q-band channels and one for both Ka-band channels. 

Upon leaving the dewar, each of the Ka-band signals travels through
waveguide to a warm amplifier. Ka2, the vertically 
polarized channel (Table~\ref{tab:centroids}), 
is amplified with a single warm amplifier with
a gain of about 50 dB and a noise figure of 3-4 dB ($T_{sys}\approx
400~$K). The signal from
Ka1, the horizontally polarized channel, was found to be 
compressed in a similar configuration so it is
passed through an amplifier ($+29$~dB), then through an
attenuator (-12 dB) and through a second amplifier (also with $+29~$dB) before
being detected with a diode. The net gain of the Ka receiver chains
is $\approx72~$dB.
 
The Q-band signals, upon leaving the dewar, are carried
via copper rectangular waveguide to 1 dB insertion loss filters (Spacek)
that filters out the low frequency tail
of the 60 GHz atmospheric oxygen line. Following the filters
the signal is amplified, sent through a band leveler (designed individually for
each channel by Pacific Millimeter) and amplified again. The
amplifiers are connected to the band leveler with a
K-connector to waveguide transition. Each amplifier
has a gain of 26-28 dB and a noise figure of $\approx 4.5$ dB. Tests of the
system with and without the band levelers are discussed in 
Monnelly (1996).  The net gain of each Q-band system is $\approx 75$~dB.

A diode detector (Millitech) at the output of the
final amplification stage of the HEMT channels converts the
incident power to a voltage. The diode sensitivity varies over the
passband and with temperature but is typically $\approx 2000~$mV/mW.
The output of the diode is connected to a low noise pre-amp with a
typical gain of 100 and roughly 2 k$\Omega$ input impedance. We tune the
DC output by selecting this impedance. Values vary by about a factor of
100 between different channels. The output of
the pre-amp is buffered and sent to the data acquisition system.

\subsection {Thermal, Mechanical, and Magnetic Considerations}

In the QMAP configuration, the HEMT amplifiers and the SIS mixers are
heat sunk to a cold plate that forms the bottom of the liquid helium
reservoir. The outer tank of the dewar is filled with liquid nitrogen. 
The vapor pressures of both the nitrogen and helium 
are held at a constant value with mechanical regulators.
The combination of pressure regulated cryogens and a balloon borne 
dewar produces a constant-temperature cold 
plate with minimal microphonics.

Due to the expense and difficulty of transporting liquid
cryogens to Cerro Toco, the TOCO receiver
is cooled with a Sumitomo cryogenic refrigerator (SRDK-408BA). 
The 40~K stage has 40~W of cooling power. The loading on the 40~K stage is
dominated by HEMT LED's, waveguide connections between the cold and warm
stages, and optical loading through the window. From the liquid
cryogen boil-off rate in the QMAP configuration, we determined that the total 
loading on this stage is $\approx 20~$W. The total loading on the  
4~K stage, where there is 1~W of cooling power, is $\approx 400$ mW. 

In Chile, diurnal variations in the optical loading led to variations in the
temperature of the SIS mixers of $\approx 400~$mK. However, for the
purposes of CMB anisotropy analysis, we used only data taken
during the night when the temperatures were stable to better than 50
mK. The temperatures were monitored continuously and the small drifts
accounted for in the analysis through a calibration model.
There was also a temperature variation of approximately 10 mK
synchronous with the refrigerator drive motor operating at 1.2
Hz. This term is asynchronous with the chopping frequency, and the 
resulting gain fluctuations are too small to affect our results 
(Section~\ref{sec-systematics}).

The backpack that houses the warm electronics mounts on the 
dewar and is thermally controlled to within $\pm 0.5$~K
over a typical night of observations. As with the cold stage, the 
temperature is monitored
continuously and the effects of drifts on gain are corrected in software.
All microwave amplifiers are heat sunk to a common
aluminum plate and their temperatures are monitored. 

The enclosure for the electronics has two levels of RF shielding and is 
filled with pieces
of microwave absorber wrapped in plastic bags that serve
the dual purpose of thermal insulator and absorber of 
stray microwave radiation (perhaps from imperfect joints).
All waveguide joints are
wrapped with an absorber to prevent leakage of radiation into the
rest of the system. We did not find any evidence for correlations
between channels due to the instrument.

If changes in RF impedance due to vibrations are synchronously modulated by
the motion of the chopping mirror, the resulting microphonic lines 
can mimic a celestial signal. Such ``microphonics'' can couple into the 
data, for example, through motion of the feeds or through 
strain in the microwave joints. If the coupling is large
and variable, it can also affect the data even if it is asynchronous 
with the chopper.

In order to minimize this coupling, the cold
head motor is vibrationally isolated from the dewar
(which is bolted to the gondola frame) with a set of compensating
flexible vacuum bellows. We also use bags of \#9 lead shot placed on the
cold head to damp vibrations. The connections between the cold head 
and the cooled electronics
are made with strips of high purity
flexible copper braid that efficiently conducts heat and 
vibrationally isolates the detectors from the refrigerator
head. 

Although microphonic levels were low at the beginning of the 1997
season, a microphonic coupling developed over the campaign that rendered
the D-band data unusable for CMB observations. The coupling
was manifest as 1.2 Hz (the cold head cycle) wings of a broad 90 Hz
line suggesting amplitude modulation of a 90 Hz vibrational line.
The source of the vibration was traced to a combination of the azimuthal
drive motor and the
chopper. It was corrected for the 1998 season by modifying these two 
systems. The electronic interference
from the azimuthal drive motor was eliminated by installing a brake.
The brake allows the motor to
be shut off during CMB observations and prevents large currents
from being drawn by the motor working to counteract wind loading. In addition,
the chopper-induced vibrations were reduced by replacing the bearings in
the chopper with flex pivots. As a result, microphonic levels in the 
TOCO98 D-band data were negligible. Microphonics were
not a problem either year in the HEMT data.  

The Josephson junctions in the SIS mixers are sensitive to magnetic
fields. Helmholz coils around the dewar were used to measure the
dependence. With the coils absent, the area was mapped with a 
Gauss meter to ensure that the AC fields from the chopper drive
and cold head motor would not contaminate the data. 
To minimize potential magnetic coupling, high magnetic permeability
material (mu-metal) was wrapped around the
outside of the chopper coils and around the cold head. 
Not only is the magnetic field negligible, but the chopper synchronous
component corresponds to values of $l$ that do not enter into 
the CMB analysis.

\begin{table*}[t]
\caption{\small Center frequency and noise bandwidths for each campaign.}
\small{
\vbox{
\tabskip 1em plus 2em minus .5em
\halign to \hsize {
     \hfil#\hfil &\hfil#\hfil &\hfil#\hfil  \cr
\noalign{\smallskip\hrule\smallskip\hrule\smallskip}
     ~~~~~~~~QMAP  & TOCO 97 & ~~~TOCO 98    \cr}
\halign to \hsize {
     #\hfil &#\hfil &#\hfil &#\hfil &#\hfil &#\hfil &#\hfil \cr
\omit & $\nu_c$ (GHz)& $\Delta\nu$ (GHz)&  $\nu_c$ (GHz) & $\Delta\nu$ (GHz)
      & $\nu_c$ (GHz)& $\Delta\nu$ \cr 
\noalign{\smallskip\hrule\smallskip}
Ka1 & $31.7\pm 0.3$ & $4.7\pm 2.2$ & $31.7\pm 0.3$& $4.7\pm 2.2$ 
    & \dots & 4.7*\tablenotemark{a} \cr
Ka2 & $30.8\pm 0.2$ & $6.2\pm 0.3$ & $30.8\pm 0.2$& $6.2\pm 0.3$
    & $32.0\pm 0.1$ & $8.8\pm 0.1$ \cr 
Q1  & $41.4\pm 0.2$ & $6.9\pm 0.3$ & $41.1\pm 0.2$& $6.9\pm 0.3$
    & \dots & 6.9* \cr
Q2  & $41.3\pm 0.2$ & $7.5\pm 0.3$ & $41.0\pm 0.2$& $7.3\pm 0.3$
    & \dots & 7.3* \cr
Q3  & $42.1\pm 0.2$ & $6.3\pm 0.3$ & $41.7\pm 0.2$& $6.3\pm 0.3$
    & $41.5\pm 0.1$ & $5.3\pm 0.1$ \cr
Q4  & $41.2\pm 0.2$ & $7.0\pm 0.3$ & $41.1\pm 0.2$& $4.6\pm 0.3$
    & $41.2\pm 0.1$ & $4.5\pm 0.1$ \cr 
D1 (USB)&\dots& \dots & $149.2\pm 0.2$ & $2.9\pm 0.2$
    & $148.7\pm 0.1$  & $2.3\pm 0.1$ \cr 
D1 (LSB)&\dots& \dots & $138.3\pm 0.2$ & $3.2\pm 0.2$
    & $139.4\pm 0.1$  & $2.6\pm 0.1$ \cr 
D1 (DSB)\tablenotemark{b}&\dots &\dots& $141.8\pm 0.7$ & $5.6\pm 0.4$~(3.1)
    & $143.7\pm 0.1$  & $4.9\pm 0.1$~(2.5) \cr 
D2 (USB)&\dots&\dots  & $148.7\pm 0.7$ & $2.5\pm 0.4$
    & $148.6\pm 0.1$  & $2.4\pm 0.1$ \cr 
D2 (LSB)&\dots& \dots & $138.9\pm 0.7$ & $ 1.7\pm 0.4$
    & $139.4\pm 0.1$  & $2.8\pm 0.1$ \cr 
D2 (DSB)&\dots& \dots & $145.4\pm 3.0$ & $3.6\pm 1.3$~(2.1)
    & $143.3\pm 0.4$  & $5.2\pm 0.1$~(2.6) \cr 
\noalign{\smallskip\hrule}
}}}
\tablenotetext{a}{The * indicates that the bandwidth is assumed from the
previous year.}
\tablenotetext{b}{The D-band channels were
not used in QMAP so no value is given. The SIS bandpasses were 
re-measured between 1997 and 1998 because they change with SIS tuning.
The numbers in parentheses following the full
RF bandwidths is the effective IF bandwidth for noise calculations.}
\label{tab:centroids}
\end{table*}

\subsection{Receiver Characteristics}

The QMAP/TOCO receiver was characterized in the lab before each
campaign, but the most relevant characterizations are done while observing.
We use the following definitions:
\begin{equation}
\nu_c = {\int{\nu g(\nu)d\nu}\over{\int g(\nu)d\nu}}~~~~~
\Delta_n\nu = {\bigg({\int g(\nu) d\nu}\bigg)^2\over\int g^2(\nu)d\nu}.
\label{eq:eff_bw}
\end{equation}
where $g$ is the receiver passband, $\nu_c$ is the effective center frequency and
$\Delta_n\nu$ is the noise bandwidth \cite{Dicke} for the
radiometer equation.\footnote{Throughout the paper, we use $\nu$ for RF 
frequencies and $f$ for audio ($<20~$kHz) frequencies.}

The noise of HEMT -based amplifiers has a $1/f$ characteristic
\cite{Jarosik96,Wollack95}. To parametrize it, we fit the power spectrum
of the detector diode output to the following form:
\begin{equation}
\tilde T = T_{sys}\sqrt{{1\over\tau\Delta_n\nu}+ f^\alpha}= 
T_{sys}\sqrt{{1\over \tau\Delta_n\nu}+ 
\biggl({\Delta G(f)\over G}\biggr)^2}
\label{eq:det_noise}
\end{equation}
where $\tilde T$ is the system sensitivity in units of K-sec$^{1/2}$,
$\tau$ is the integration time, and
$\Delta G/G$ is the fractional gain fluctuation that gives rise to
the $1/f$ form. When the bandwidth is very large, the gain fluctuations
dominate as shown by Wollack \& Pospieszalski (1998). 
Note that a $1/f$ noise spectrum corresponds to $\alpha=-1$ in the
variance (``power'') of the detector output. We define the $1/f$ knee,
$f_{knee}$, to be the frequency where the power spectrum (square of
equation~\ref{eq:det_noise}) increases by
a factor of two over the value at high frequencies. 

The instrument bandpasses, $g(\nu)$, are
measured in the lab for each channel. Table~\ref{tab:centroids} lists
the center frequencies and effective noise bandwidths for each channel
for each observing campaign as calculated from equation~\ref{eq:eff_bw}. 

The SIS is operated in double side band (DSB) mode. By convention we use
the intermediate frequency (IF) noise
bandwidth in the radiometer equation and report ``double sideband noise
temperatures,'' $T_{DSB}$, because our source fills both 
RF bands. If a source fills
just one RF band, a ``single sideband noise,'' $T_{SSB}$, is reported. In
an ideal system, $T_{DSB}=T_{SSB}/2$. We use the mean of the USB and
LSB bandwidths for the noise bandwidth in the radiometer equation.
A full calculation of the noise includes contributions from the mixer
and the IF amplifier \cite{Blundell92,Kerr97}; for our purposes we treat
these as lumped elements. 

\subsubsection {SIS Sensitivity}
Measurements of the SIS sensitivity have been made in
several configurations as shown in Table~\ref{tab:sis_sensitivity}.
The values of $T_{\rm rec}$ from laboratory measurements 
are better than those made in the field. To investigate this
discrepancy, we built an external cold load
that could be bolted onto the front of the receiver to mimic the sky.
We found that it was possible to approximately reproduce the system temperatures
measured with the internal load using an external load provided that the
tuning parameters were readjusted. This retuning compensates primarily
for the change in temperature of the mixer. In the 1998 season, the
combination of thermal loading and increasingly poor refrigerator performance 
led to an increase in the SIS temperatures and a temperature
distribution different to that in the lab. Even though the system was
tuned in the field, the lab performance was not duplicated.

The SIS system noise exhibits a $1/f$ characteristic presumably due to
the C-band HEMT, though this has not been verified.
The $1/f$ noise is parametrized following
equation~\ref{eq:det_noise} in the last two lines of
Table~\ref{tab:sis_sensitivity}.

\begin{table*}[t]
\caption{SIS system parameters}
\small{
\vbox{
\tabskip 1em plus 2em minus .5em
\halign to \hsize {
     #\hfil &#\hfil &#\hfil &#\hfil &#\hfil \cr
\noalign{\smallskip\hrule\smallskip\hrule\smallskip}
& D1 (lab int.)& D1 (field)& (lab int.)& (field) \cr
\noalign{\smallskip\hrule\smallskip}
${\rm{T}_{\rm phys} ({\rm K})}$\tablenotemark{a} & 4.3 & 4.9 & 4.4 & 4.9 \cr
${V_B({\rm mV})}$\tablenotemark{b} &13.4&13.8&14.2&14.6 \cr
%
%
$\tilde{T}_{DSB}({{\rm mK}\sqrt s})$\tablenotemark{c}&0.6&1.3&0.7&1.2 \cr
$T_{\rm sys}$\tablenotemark{d} &30&65&$\cdots$&61 \cr
$T_{\rm rec}$\tablenotemark{e}&26&48&$\cdots$&44 \cr
$T_{\rm rec}({\rm y-factor})$\tablenotemark{f}&27&$\cdots$&35&$\cdots$ \cr
$\alpha$        & $\cdots$ & -1.0 & $\cdots$  & -0.8 \cr
$f_{knee}$ (Hz) & $\cdots$ & 18   &  $\cdots$ & 12 \cr
\noalign{\smallskip\hrule\vskip1pt\hrule\smallskip}
}}}
\label{tab:sis_sensitivity}
\tablenotetext{a}{The physical temperature of the SIS body. The uncertainty is
$\approx\pm 0.2$~K. At $T>5~$K the SIS sensitivity is markedly reduced.}
\tablenotetext{b}{The optimal SIS bias voltage,
$V_B$, across the 6 SIS junctions.}
\tablenotetext{c}{The total power DSB sensitivity computed from the
noise power spectrum at 200 Hz, where atmospheric fluctuations are
negligible, and the responsivity. These are for a
Rayleigh-Jeans source. In the field, the atmosphere and telescope
contribute $\approx17~$K. In the lab, the load contributes
$\approx4$~K. The loss from the feeds is measured to be negligible.}
\tablenotetext{d}{The Rayleigh-Jeans system temperature computed from measured
sensitivity and the noise bandwidth.}
\tablenotetext{e}{The Rayleigh-Jeans receiver temperature computed from $T_{sys}$.}
\tablenotetext{f}{The Rayleigh-Jeans receiver temperature measured with a
variable temperature cryogenic load.}
\end{table*}

\begin{table*}
\caption{\small HEMT amplifier system parameters}
\small{
\vbox{
\tabskip 1em plus 2em minus .5em
\halign to \hsize {
     \hfil#\hfil &\hfil#\hfil &\hfil#\hfil  \cr
\noalign{\smallskip\hrule\smallskip\hrule\smallskip}
     ~~~~~~~~~QMAP\tablenotemark{a}  ~~~~~~~~~& TOCO 97\tablenotemark{b} 
          & ~~~~~~~~~~~~TOCO 98 \cr}
\halign to \hsize {
 \hfil#\hfil &\hfil#\hfil &\hfil#\hfil &\hfil#\hfil &\hfil#\hfil 
&\hfil#\hfil &\hfil#\hfil &\hfil#\hfil &\hfil#\hfil &\hfil#\hfil
    &\hfil#\hfil  &\hfil#\hfil  &\hfil#\hfil  \cr
  & $T_{\rm sys}$ & $\tilde{T}$ & $\alpha$      & $f_{knee}$ 
  && $T_{\rm sys}$ & $\tilde{T}$ && $T_{\rm sys}$ & $\tilde{T}$ & 
    $\alpha$ & $f_{knee}$\cr
  & K & ${\rm mK}\sqrt{\rm s}$ &&  Hz 
  && K & ${\rm mK}\sqrt{\rm s}$ && K & ${\rm mK}\sqrt{\rm s}$ && Hz \cr
\hline\hline
Ka1& 22 & (0.40)~0.36 & -1.2   & 62    && 89 & 1.3 && 
          162 & (2.6)~2.4    & -0.90  & 14 \cr
Ka2& 23 & (0.42)~0.32  & -0.92  & 141   && 63 & 0.8 &&
           59 & (0.9)~0.62   & -0.85  & 95\cr
Q1 & 17 & 0.21  &\dots   & \dots && 91 & 1.1 &&
           84 & (1.2)~1.0    & -0.84  & 64\cr
Q2 & 22 & (0.60)~0.25      & -0.82  & 686 && 145 & 1.7 && 
          114 & (3.3)~1.3    & -1.57  & 168\cr
Q3 & 155 & (2.1~)~2.0   & -0.78  & 20 && 63 & 0.8 &&
          80  & (0.9)~0.72   & -0.84  & 37 \cr
Q4 & 53 & (0.85)~0.64   &-0.71   & 67 && 156 & 2.3 && 
          88  & (1.6)~1.3    &-0.88    &42\cr
\noalign{\smallskip\hrule}
}}}
\tablenotetext{a}{From the first QMAP flight as shown in \cite{dev98}.
Q1 did not work during the flight so we report the lab
measurements \cite{Monnelly96}.
The four entries for each campaign correspond to the system
temperature, the measured system sensitivity at 100 Hz (in parentheses)
and very high frequencies, the gain fluctuation exponent, 
and the $1/f$ knee. In the
fits to the power spectra, $f<5~$Hz is not included. Because of the
$1/f$ HEMT characteristics, $\tilde T= T_{sys}/\sqrt{\Delta_n\nu}$ is always 
smaller than the measured value of $\tilde T$ at 100 Hz.  
With the centroids in Table~\ref{tab:centroids}, the full noise spectrum 
may be recovered.}
\tablenotetext{b}{From measurements in the field at 100 Hz.}
\label{tab:hemt_sensitivity}
\end{table*}

\subsubsection {HEMT Sensitivity}

The same set of six HEMT amplifiers were used for the two QMAP flights
and the two observing seasons of the TOCO experiment; four of this
set were used for the Saskatoon measurement. They were 
tested in the lab before each set of observations and the sensitivities were
analyzed for each data set (Table~\ref{tab:hemt_sensitivity}). There is
evidence of degradation in the HEMT performance between the
QMAP and TOCO campaigns above that which is expected due to the difference in
body temperature \cite{posp89}.
Generally HEMT system noise is expected to increase roughly one Kelvin
for each Kelvin of increased ambient temperature. 
We suspect slow deterioration in the mechanics of the microwave/bias
structures over the hundreds of cycles and sometimes rough handling
these amplifiers experienced.
The chips were unpassivated InP so there may some deterioration in the 
chip performance though this has not been confirmed.

%
%
\section{Optics}
The telescope optics are similar to those used in the Saskatoon
experiment~\cite{Wollack97}. Corrugated feeds
underilluminate a 0.86~m
primary mirror which in turn underilluminates a flat (1.8~m$\times$1.2~m)
chopping mirror (chopper). Each of the eight channels detects a single
mode of a diffraction limited beam. The chopper is a resonant, computer
controlled mirror, that scans in the azimuthal direction while the
rest of the optics remain fixed in azimuth and elevation as the sky
rotates through the beams.  The
telescope sits inside an aluminum ground
screen which is fixed with respect to the primary mirror, the receiver, and
the chopper mount. 

\begin{table*}
\caption{Design Parameters for All Feeds}   
\small{
\vbox{
\tabskip 1em plus 2em minus .5em
\halign to \hsize {
     #\hfil &#\hfil &#\hfil &#\hfil &#\hfil \cr
\noalign{\smallskip\hrule\smallskip\hrule\smallskip}
& Ka&Q1/2&Q3/4&D \cr
\noalign{\smallskip\hrule\smallskip}
Semiflare angle $\theta_o$ (deg)& 6& 4.4& 5.4& 5.4 \cr
Skyward aperture diameter $d_h$(cm)& 4.2 & 2.0 & 2.1 & 0.89 \cr
OMT aperture diameter (cm)& 0.833 & 0.650 & 0.650 & 0.173 \cr
Beamwidth $\theta_{\rm beam}^{\rm FWHM}~$(deg)& 18 &
        18 & 16 & 17\cr
Phase error $\Delta$ \cite{macathomas}&0.11&0.065&0.051&0.011\cr
VSWR & 1.03 & 1.03 & 1.03 & 1.05\cr
Forward gain (dBi) & 20.9 & 20.6 & 21.4 & 20.8\cr
Number of corrugations& 71& 67& 85& 82 \cr
Length (cm)& 19.3& 12.0& 14.5& 3.96 \cr
\noalign{\smallskip\hrule\vskip1pt\hrule\smallskip}
}}}
\label{tab:feeds}
\end{table*}

\subsection {The focal plane}
\label{sec:focal_plane}
Conical corrugated feed horns receive radiation from the sky and
transform the incident fields so they may propagate through waveguide. 
All of our feeds were fabricated by Custom Microwave from
electroformed copper over an aluminum mandrel. They are gold coated 
to stabilize the surface.  The general electromagnetic design follows the
guidelines in Clarricoats and Olver (1984) and MacA Thomas (1978). 
The throat section, where the corrugations 
adiabatically transform from $\lambda/2$ depth to $\lambda/4$ depth
as the hybrid mode detaches from the feed wall, was designed
by Wollack (1994) following the work of James (1982)
\footnote{In Ka band, the depths of grooves 1-10 are:
0.411, 0.368, 0.351, 0.335, 0.323, 0.312, 0.302, 0.292, 0.282, and 0.274
cm. This feed was designed by Ed Wollack.}   
The VSWR is less than 1.05 across the waveguide band and the loss
in the feed is negligible. We model the feeds with a commercial
code \cite{YRS}, {\tt CCHRN}, that solves for the full electromagnetic
field that propagates in the feed. Table~\ref{tab:feeds} summarizes
parameters of the feed
horns used in the QMAP and TOCO experiments. 

The Ka and Q1/2 feeds were also used in the Saskatoon experiments
where they had three concentric choke grooves on 
the flange that defined the skyward aperture. It was 
believed that these would reduce ``edge currents'' resulting in a
cleaner beam. To move the feeds as close together as possible in the 
focal plane, we removed those grooves and saw no degradation in
performance. 

\begin{figure*}
\plottwo{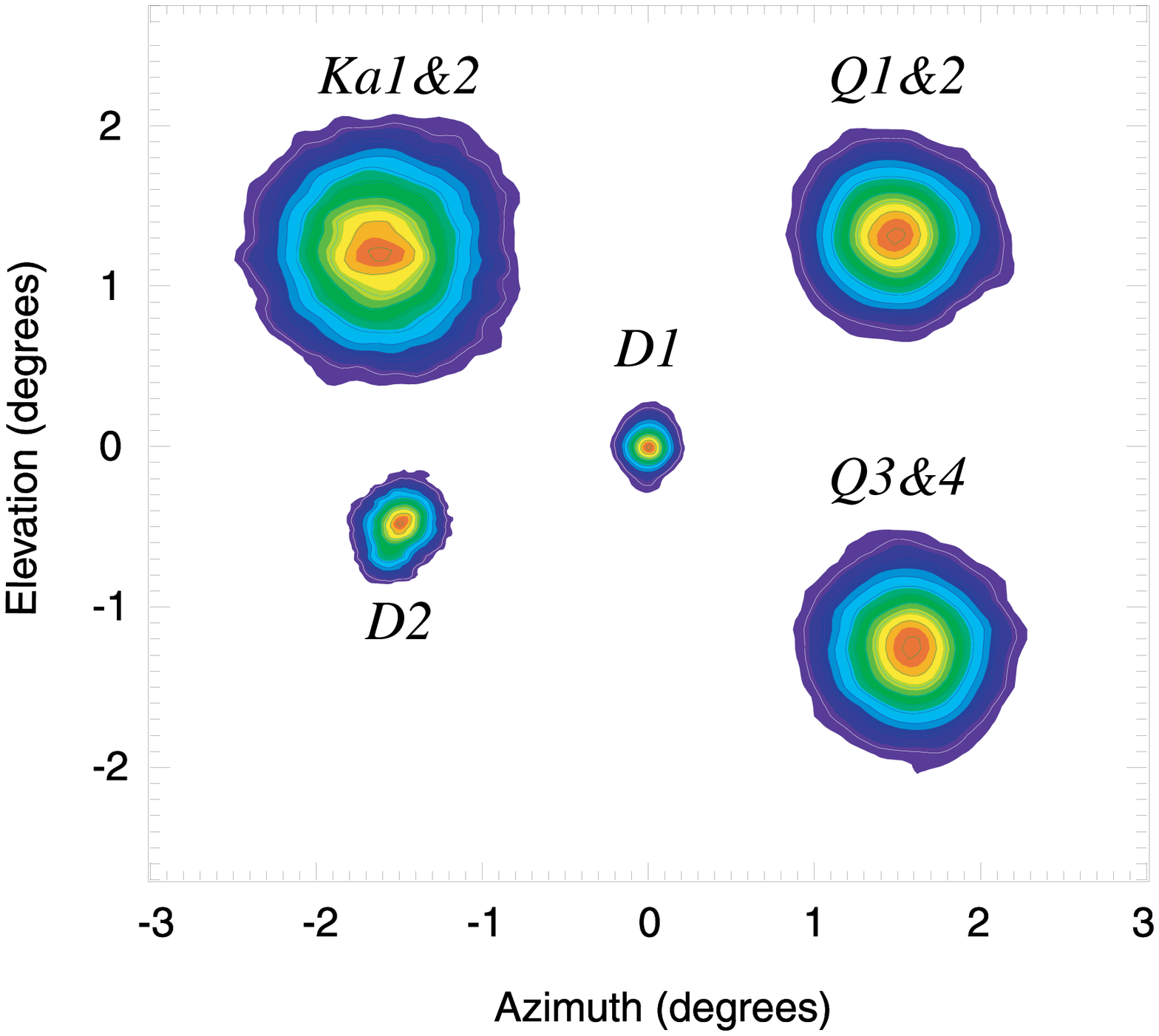}{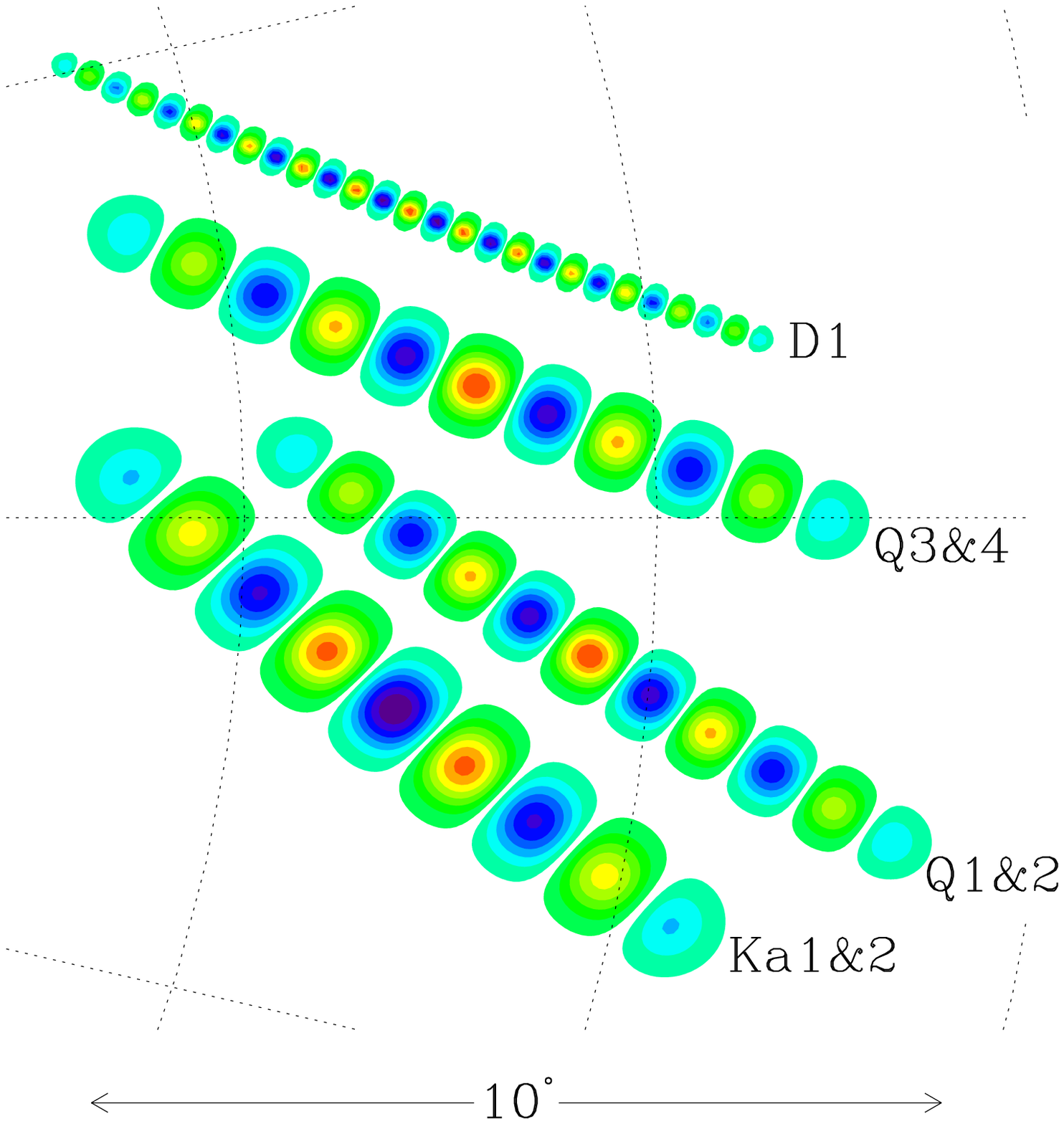}
\caption{Beam map of focal plane (left) and the synthesized beams (right). 
{\it Left:} Units are degrees on the sky from the 
center of the focal plane. Each beam is normalized with the contours
representing $10\%$ in amplitude. The physical
separations between the Ka1/2 horn and the Q1/2 feed is $\approx 5$
cm. D2 is offset from the focal plane center by 2.9~cm.
The distance between the Q1/2 and Q3/4 feeds is also $\approx 5$ cm.
This is also a picture of the up/down reflection of the feeds in the
focal plane when looking into the dewar. {\it Right:}
The synthesized beams for TOCO97 as discussed in
Section~\ref{sec-demod}. If the Ka
primary beam on the left, for example, is weighted by positive and negative
numbers as it moves across the sky, one obtains the synthesized
Ka pattern on the right. The contours indicate alternate positive and
negative lobes. Shown are the Ka 9-pt, Q 11-pt,
and D 27-pt D beams. From this picture, it is clear that the synthesized beam
is sensitive to only a narrow band of spatial frequencies. The central
dashed line corresponds to RA=0 (Fig.~\ref{fig:skyscan}) though
the synthesized beam location as shown is arbitrary. In the full
analysis, the beam is smoothed in right ascension.}
\label{fig:1focal_plane}
\end{figure*}

Figure~\ref{fig:1focal_plane} shows a map of the beam pattern made by
observing Jupiter. The feed horns are arranged so that the
D-band beams with
$0.2\deg$ resolution are placed as close to the center of the focal
plane as possible. 
Beam parameters for each channel
have been calculated and measured for
each campaign as shown in Table~\ref{tab:beams}. 

\begin{table*}
\caption{Beam Parameters}
\footnotesize{
\vbox{
\tabskip 1em plus 2em minus .5em
\halign to \hsize {
     #\hfil &#\hfil &#\hfil &#\hfil &#\hfil &#\hfil &#\hfil &#\hfil &#\hfil \cr
\noalign{\smallskip\hrule\smallskip\hrule\smallskip}
  Campaign & Ka1 & Ka2 & Q1 &  Q2 & Q3
           & Q4 & D1 & D2\cr
\noalign{\smallskip\hrule\smallskip}
 Predicted   \cr
 $\Omega_{A}$\tablenotemark{a}~($10^{-4}$~sr)
  &2.76&2.76&1.53&1.53&1.69&1.69&0.124& $\cdots$\cr 
 $\theta_{FWHM}^{az}$\tablenotemark{b} ~(deg)&
   0.905&0.905&0.663&0.663&0.702&0.702&0.190&$\cdots$\cr
 $\theta_{FWHM}^{el}$\tablenotemark{b} ~(deg)
   &0.888&0.888&0.661&.661&0.683&0.683&0.192&$\cdots$\cr 
 Azimuth\tablenotemark{c} ~(deg)&203.13&203.13&206.75&206.75&206.70
&206.70&205.00&$\cdots$\cr 
 Elevation (deg)&41.8&41.8&41.8&41.8&39.2&39.2&40.66&$\cdots$    \cr
Polarization &$\leftrightarrow$&$\updownarrow$&
        $\leftrightarrow$&$\updownarrow$&$\leftrightarrow$&
        $\updownarrow$&$\leftrightarrow$&$\updownarrow$  \cr
Pri ET (dB)\tablenotemark{d}&-21&-21& -19& -19&-20&-20&-22&$\cdots$    \cr
Chop ET (dB) &-47&-47&-48&-48&-49&-49&-54&$\cdots$    \cr 
\noalign{\smallskip}
 QMAP96a  \cr
$\Omega_{A}$ ($10^{-4}$~sr)&2.83&2.83&1.58&1.58&1.62&1.62& $\cdots$&$\cdots$\cr
$\theta_{FWHM}^{Maj}$
   (deg)&0.931&0.931&0.694&0.694&0.700&0.700&$\cdots$&$\cdots$\cr 
$\theta_{FWHM}^{Min}$ (deg)
   &0.882&0.882&0.658&0.658&0.668&0.668&$\cdots$&$\cdots$\cr 
Cross-El (deg) &-1.35&-1.35&+1.35&+1.35&+1.35&+1.35&$\cdots$&$\cdots$\cr
\noalign{\smallskip}
 QMAP96b   \cr
$\Omega_{A}$ ($10^{-4}$~sr)&2.67&2.67&1.43&1.43&1.75&1.75&$\cdots$&$\cdots$\cr
$\theta_{FWHM}^{Maj}$
   (deg)&0.932&0.932&0.674&0.674&0.730&0.730&$\cdots$&$\cdots$\cr
$\theta_{FWHM}^{Min}$ (deg) &
    0.831&0.831&0.616&0.616&0.694&0.694&$\cdots$&$\cdots$\cr 
Cross-El (deg)&-1.35&-1.35&+1.35&+1.35&+1.35&+1.35&$\cdots$&$\cdots$\cr
Elevation (deg) &41.36&41.36&41.37&41.37&38.79&38.79&$\cdots$&$\cdots$\cr
\noalign{\smallskip}
 TOCO97\tablenotemark{e}   \cr
$\Omega_{A}$ ($10^{-4}$~sr)&
    2.76&2.73&1.63&1.74&1.76&1.78&0.183&0.323\cr 
 $\sigma_\Omega$\tablenotemark{g}& 6\% & 6.5\% & 4\% & 8\% &
 4\% & 6.0\% & 5\% & 6\%\cr
$\theta_{FWHM}^{az}$ (deg) & 0.881&0.871&0.711&0.744&0.718&0.716&0.225&0.306\cr
$\theta_{FWHM}^{el}$ (deg) &0.909&0.909&0.664&0.676&0.711&0.721&0.236&0.306\cr
Azimuth (deg)& 203.13& 203.13&206.75&206.75&206.70&206.70&205.00&$\cdots$\cr
Elevation (deg) &41.75&41.75&41.85&41.85&39.25&39.25&40.44&39.93\cr
\noalign{\smallskip}
 TOCO98    \cr
 $\Omega_{A}$ ($10^{-4}$~sr)&3.00&3.00&1.52&1.60&1.76&1.80&
0.136\tablenotemark{f}&0.292\cr 
 $\sigma_\Omega$&$8\%$&$8\%$&$9.8\%$&$18\%$&
 $8.4\%$&$10\%$& $5.5\%$&$5\%$ \cr
 $\theta_{FWHM}^{az}$ (deg)
   &0.860&0.914&0.666&0.669&0.692&0.688&0.201&0.293\cr
 $\theta_{FWHM}^{el}$ (deg) &
   0.907&0.918&0.659&0.681&0.732&0.754&0.194&0.286    \cr 
 Azimuth (deg) &205.67&205.67&209.16&209.16&209.06&209.06&207.47&205.73 \cr 
 Elevation (deg) &42.05&42.05&42.03&42.03&39.48&39.48&40.63&40.13    \cr 
\noalign{\smallskip\hrule\vskip1pt\hrule\smallskip}
}}}
\label{tab:beams}
\tablenotetext{a}{Solid angle of beam.} 
\tablenotetext{b}{The full width at half maximum in the azimuthal
and elevation direction. The beam is not symmetric due to smearing 
in the azimuth direction and due to the placement in the focal
plane. For QMAP, we give the major and minor axes of the best fit 
ellipsoidal Gaussian.} 
\tablenotetext{c}{The azimuth and elevation of the beams. For QMAP, we
give the relative position in the focal plane. For TOCO98, the pointing
was slightly different.}
\tablenotetext{d}{The edge taper is the ratio of radiant power in the
center of the optic to that on the edge expressed in dB. This is
determined from a full calculation of the current distribution
on each optical element using the DADRA program \cite{YRS}.}
\tablenotetext{e}{The chopper was tilted around its normal for TOCO97
and so the beam moved in elevation by $0.116\sin(\theta_{az-chop})$~deg. 
There was no tilt in TOCO98. }
\tablenotetext{f}{The prediction is for the 1998 configuration. D band
is also the most sensitive to slight changes in alignment.}
\tablenotetext{g}{The {\it rms} in the solid angle during the
campaign rounded to the nearest 0.5\%.}
\end{table*}

Uncertainty in the beam solid angle leads to calibration uncertainty.
Confidence in our understanding of the beams has been built by observing
a variety of sources over two years and comparing those results with
detailed models. QMAP mapped the beams using Cas-A
and Saturn; TOCO mapped Jupiter more than 70
times. Each set of observations was independently fit. 
Variations from year to year are due to different alignments of the
parabola and feeds. 

\subsection {Optical components}
\label{mirrors_groundscreens}
The primary mirror is a simple offset parabola. In the notation
of  Wollack {\it et al.} (1993), the offset angle is 
$\Psi_P=48^{\circ}$, and the focal length is 0.9 m. The dewar views the
parabola at an angle of $\Psi_D=7^{\circ}$ thus the zenith angle of the
optical axis is $90^{\circ}-\Psi_P+\Psi_D=49^{\circ}$. The diameter of the
parabola in projection is 0.86 m. The {\it rms} surface 
tolerance is $\approx15~\mu$m.

The chopper design follows that described in Wollack (1997) with a few
modifications. It is smaller and lighter than the Saskatoon chopper,
measuring 1.8 by 1.2 m, and
is driven at the resonant frequency of a flat steel spring plate
(85~cm by 8.62~cm by 0.167~cm for TOCO98) attached at its middle to the
chopper mount. The surface tolerance of the plate is $\approx30~\mu$m. 
The thickness of the spring plate is tuned for each campaign.
The resonant system produces a sinusoidal sweeping
pattern in azimuth that requires minimal drive power and produces minimal
vibrations on the mount. There is no reaction bar on the QMAP/TOCO chopper. 

The chopper position is sensed 80 times per chop with a set of redundant LVDTs
(Linear-Variable-Differential-Transformers), one on each side of the
mirror. Errors in the sinusoidal chopper
motion due to wind loading or instrument glitches are measured. 
The temperature of the chopper is monitored at all times using
nine thermometers placed at various locations on the flat.
For TOCO, when the {\it rms} chopper position over a single chop deviates from
the expected position over the chop by more that $0\deg.015$ the data are rejected.
The chopper zero point position in azimuth was measured as a function of
time in the field and found to change less than $0\deg.01$.
For QMAP, no data cuts were made based on the chopper motion.

\section {Offsets and Sidelobes}
\label{sec-offsets}

Emission from the instrument produces signals that can potentially 
complicate the measurement\footnote{The Sun is another source. 
For TOCO, the Sun travels
overhead and so cannot be completely blocked at all times. However, for CMB
work we use data only from when the Sun is fully blocked from the
optics. QMAP flew at night. Lunar emission was not seen in either experiment.}.
We generically call these signals ``offsets'' as they are fairly constant
over long periods and would be present in the absence of a celestial
signal. Examples are shown in Figure \ref{fig:obs_offset}.
Outside of the data selection based on weather, most of the
analysis effort goes into ensuring that offsets do not contaminate
the final results.
This section addresses the known offsets and sets upper limits
on their magnitudes for both the case of 
when the source is chopper-modulated (radiation
which enters the detectors after being affected by the chopper) and
unmodulated (radiation which enters the detectors directly).
The offsets from modulated emission can, in principle, occur at any harmonic
of the chopper frequency\footnote{We use the term ``harmonic'' to refer
to the spatial frequencies of the scan pattern as discussed in
Section~\ref{sec-demod}.}. In practice, they are predominantly
at the lowest spatial frequency harmonics. 

In the following we focus on TOCO because the long
observing campaigns required careful monitoring of the offsets. 
In QMAP, because of the short durations of the flights, the offsets were
stable. In the mapmaking analysis, the offsets were projected out of
the CMB data using a technique described in
de Oliveira-Costa {\it et al.} (1998).

\begin{table*}
\caption{Estimated Contributions to the Offsets}
\small{
\vbox{
\tabskip 1em plus 2em minus .5em
\halign to \hsize {
     #\hfil& #\hfil &#\hfil &#\hfil &\hfil#\hfil \cr
\noalign{\smallskip\hrule\smallskip\hrule\smallskip}
  Contributions to the Signal & D-Band & Q-Band & Ka-Band  &Modulated\cr
\noalign{\smallskip\hrule\smallskip}
 Earth emission (directly into feed)\tablenotemark{a}& 75 $\mu$K & 300 $\mu$K & 400
	$\mu$K& No \cr
 Earth emission (in chopper sidelobes)\tablenotemark{b}&$<50~\mu$K& $<200~\mu$K &
        $<250~\mu$K& No \cr
 Ground screen emission (directly into feed)&5 K& 6 K& 7 K& No\cr
 Ground screen emission (in chopper sidelobes)&80 mK&80 mK &80 mK& No \cr
 Cavity emission, vertically polarized & 3 mK& 13 mK& 15 mK& No \cr
Total loading from optics & 5 K & 6 K  & 7 K  & \cr
\cr
 Total ground screen emission\tablenotemark{c}&1.5 mK&1.5 mK &1.5 mK & Yes\cr
 Cavity emission (vertically polarized)\tablenotemark{e}~~~
      & 0.3~mK& 1.3~mK& 1.5~mK& Yes \cr
 Polarized chopper emission\tablenotemark{d}& 3.5 mK& 1.8 mK
  & 1.6 mK & {\bf Yes}\cr
 $\delta T_{\parallel}$ mirror misalignment in atmosphere& 3 mK& 1.7 mK&
1.7 mK & {\bf Yes}\cr
$\delta T_{\perp}$ mirror misalignment in atmosphere& $20$ $\mu$K& $10$ $\mu$K&
	$10$ $\mu$K & {\bf Yes} \cr
 Feed rotation ($1\deg$) & 3~mK& 0.8~mK& 0.6~mK & {\bf Yes}\cr
 Feed rotation ($4\deg$) & 10~mK& 3~mK& 2~mK & {\bf Yes}\cr
\noalign{\smallskip\hrule\vskip1pt\hrule\smallskip}
}}}
\label{tab:offsets_summary}
\tablenotetext{a}{Results of the formal calculation times three to
account for modeling errors.}
\tablenotetext{b}{Includes a factor to account for near field effects.}
\tablenotetext{c}{Due to a hypothetical 
temperature gradient across the ground screen.}
\tablenotetext{d}{Magnitude of the largest component of polarized
emission. The bold ``{\bf Yes}'' indicates that this quantity is computed to
10\% accuracy.}
\tablenotetext{e}{We estimate that the effective area of the cavity 
between the chopper and the outer feed baffle is modulated by 10\%.} 
\end{table*}

\subsection {Earth Emission Offset}
\label{sec-earth_emission}
Radiation from the Earth can diffract over the front edge of the ground
screen and enter the receivers. The temperature of the ground as seen by
the detectors is

\begin{equation}
T_{A}\approx{g_{\rm feed}(\theta)\over{4\pi}}\bigg[{D^2\over r}\bigg]T_E\Omega _{E}
~~~~~~~~~{\rm with}
\nonumber
\end{equation}
\begin{equation}
D=-{\sqrt{\lambda}\over 4\pi }\bigg[
{1\over {\rm cos}((\gamma-\alpha)/2)}\pm {1\over 
{\rm sin}((\gamma+\alpha)/2)}\bigg]
\label{eq:T_det}
\end{equation}
where $T_E$ is the physical temperature of the Earth ($\approx 273$ K);
$g_{\rm feed}(\theta)$ is the gain of the feed as defined by 
$G_{max}P_n(\theta)$ where $G_{max}$ is the forward gain and  
$P_n(\theta)$ is the normalized beam pattern;
$\Omega _{E}$ is the solid angle of the Earth subtended by the
telescope rim; $r$ is the distance from the horn to the top 
front edge of the baffle; $\gamma$ and $\alpha$ are the diffraction
angles; and 
D is the diffraction coefficient~\cite{Keller}. The positive sign
in D is for the E-field perpendicular to the edge and the
negative sign is for the E-field parallel to the edge. 

From integrating equation~\ref{eq:T_det} over the geometry of the
ground screen and the feed pattern, we find the diffracted power in
D-band into the feed $T_{\rm A}\approx 25~\mu$K. The
front baffle, which is in the far field of the feed, 
has a ``Keller flare'' that reduces the diffracted
power over that from the sharp edge we assumed in the calculation.
Based on our experience, calculations of this type involving complicated 
geometries are accurate to roughly a factor of three. We have included
this factor in Table~\ref{tab:offsets_summary} where results are summarized.

We estimate the modulated contribution the same way as
above but with $g_{\rm feed}(\theta)$ replaced by $g_{\rm
beam}(\theta)$, the far-field gain of the main beam rather than the gain of
the horn. The calculated power diffracted into the chopper sidelobes 
is $\approx5~\mu$K. The front baffle is in the near field of the main
beam and so the true values could be up to an order of magnitude larger. 
If the temperature of the ground on
either side of the telescope were to differ by 10 K then 
the offset produced would be $< 1~\mu$K. Similarly, any modulation
of diffracted power from variations along the top of the ground screen
are small.
From these estimates we conclude that emission from the Earth does
not contribute to the microwave signal.

\subsection {Ground Screen Emission Offset}

The antenna temperature from emission of the baffles as seen by the receiver is
\begin{equation}
T_{A}\approx{g_{\rm feed}(\theta)\over{4\pi}}\epsilon_{baffle}T_{\rm baffle}\Omega
_{\rm baffle}\approx 5{\rm K}.
\label{eq:T_det_baffle}
\end{equation}
where $\epsilon_{baffle}$ is the emissivity of the
baffle and $T_{\rm baffle}$ is its physical temperature.
Along with the atmosphere, the direct emission from the ground screens 
is the largest radiation loading.
As with the diffracted ground emission, we
ask what portion of this signal is modulated. 
We find that the temperature of the baffle in the main beam, after
reflecting off the chopper, is $\approx
80$ mK. A temperature differential of 10 K on either side of the baffle would
then translate to an observed offset of $\approx 1.5$ mK in the lowest
harmonics. It is also possible to get modulated emission because the 
polarized emission is a function of the angle of the beam with respect
to the ground screen. The geometry of the enclosure is complicated, but 
the angle of the ground screen implies that the emission will be greatest in the
vertical polarization. However, the modulation will be greatest
for the horizontal component. This term is similar in character to the
polarized emission from the chopper but is an order of magnitude smaller.

\subsection{Cavity Emission Offset}
\label{sec-cavity_emission}

The cavity behind the chopping mirror (Fig.~\ref{fig:dewar_shells})and
outer feed baffle is effectively black. 
Radiation from this cavity reaches the receiver: a) by traveling
over the front edge of the outer feed baffle and back into the feeds
or b) through shallow angle diffraction over the outer feed baffle
onto the parabola and then
reflecting back to the feeds. This emission
is modulated as the chopping mirror sweeps back and forth,
changing the size and shape of the opening to the emitting cavity.
Improved shielding of this cavity between the two QMAP
flights reduced the offset at a given angle of the chopper by 1.8~mK or
50\% in Ka band \cite{Herbig98}.  

The contribution from radiation traveling over the feed baffle and back
into the feeds may be estimated with equation~\ref{eq:T_det}.
For D-band we find that $T_{A}\approx3~$mK for vertically polarized
radiation and $T_{A}\approx0.1~$mK for horizontally polarized
radiation. 

The contribution from the shallow angle diffraction is difficult
to compute to even order-of-magnitude accuracy because it is so 
critically dependent on the geometry. We note, though, that this term 
can be of millikelvin magnitude and horizontally polarized.

\subsection {Chopper Polarized Emission Offset}\label{polarized_emission}
The emission from the chopping plate is polarized and depends
on the angle of the mirror \cite{landl60,Wollack93,cortig94}.
Therefore, as the chopping
mirror scans across the sky the plate emissivity, as viewed by a feed,
changes with the position of the chopper. 
The parallel and perpendicular
emissivities are:
\begin{equation}
\epsilon_{\parallel}\approx\epsilon_0/{\rm cos}(\theta _i)~~\&~~
\epsilon_{\perp}\approx\epsilon_0{\rm cos}(\theta _i),
\label{eq:e_perp}
\end{equation}
where $\epsilon_0$ is the emissivity at normal incidence and the 
incident angle of radiation is
\begin{equation}
\theta _i(\phi_c)={\rm cos}^{-1}[-\hat{k}_i\cdot\hat{n}(\phi_c)],
\label{eq:rad_inc}
\end{equation}
where $\hat{k}_i$ is the propagation vector for the incoming ray,
$\hat{n}$ is the normal to the plate, and $\phi_c$ is the chopper angle
in the azimuthal direction. The brightness temperature of the plate is 
\begin{equation}
T_{emit}\approx
T_{chop}[\epsilon_{\parallel}|\tilde{E}_\parallel(\theta_i)|^2+
\epsilon_{\perp}|\tilde{E}_\perp(\theta_i)|^2],
\label{eq:plate_brightness}
\end{equation}
where $\tilde{E}_\parallel$ and $\tilde{E}_\perp$ are the parallel and
perpendicular electric field projections onto the normal to the chopping
mirror and $T_{chop}$ is the physical temperature of the 
chopper. For the D-band, the peak-to-peak calculated offset is
$\approx 3.5$~mK. The dashed curves in Figure~\ref{fig:obs_offset}
show the polarized emission offset calculated for each of the
channels. 

In the beam synthesis, the net signal for each chopper position is multiplied 
by the corresponding element in the synthesis vector (Section~\ref{sec-demod}). 
The offset, since it is an additive signal, is multiplied by the same vectors.
It is evident from  Figure~\ref{fig:obs_offset} that the offset
in the synthesized signal, or ``synthesized offset,'' will be 
larger for the smaller harmonics.
The polarization directions were chosen to minimize
the synthesized offset for the 3-pt beam as no science data
is expected from 2-pt beam.

\subsection {Atmospheric Offsets}
\label{sec-atmospheric_offsets}

When the telescope is properly aligned, the chopper scans horizontally
through the atmosphere and the atmospheric emission temperature at all
portions of the chop is the same. If either the chopping mirror or the
entire base is misaligned with respect to the horizontal, offsets are 
produced \cite{Wollack97} according to 
\begin{equation}
T_{atm}=<T_{atm}>+{\partial T_{atm}\over
\partial\psi}\bigg[\delta\psi_{\parallel}+\delta\psi_{\perp}\bigg]
\label{eq:atm_offsets}
\end{equation}
where 
\begin{equation}
{\partial T_{atm}\over\partial\psi}=T_z {\rm tan}(\theta_z){\rm sec}(\theta_z)
\label{eq:atm_grad}
\end{equation}
is the gradient in the atmospheric temperature, $T_z$ is the zenith
temperature, and $\theta_z$ is the fiducial zenith angle of the beam. 
If the chopper
is misaligned due to a rotation about an axis parallel to the chopper
normal, the sky signal is changed by
\begin{equation}
\delta\psi_{\parallel}\approx
2\delta\theta_{\parallel}{\rm sin}(\theta_z){\rm sin}(\phi_c)
\label{eq:deltat_parallel}
\end{equation}
where $\phi_c$ is the azimuthal chopper angle. If the chopper
is misaligned due to a rotation about an axis perpendicular to the
chopper normal, the sky signal is changed by
\begin{equation}
\delta\psi_{\perp}\approx
-2\delta\theta_{\perp}{\rm cos}(\theta_z){\rm cos}(\phi_c).
\label{eq:deltat_perp}
\end{equation} 
Measurements of chopper and base tilt put a limit on the measured value
of $\delta \theta_{\parallel}$ and $\delta \theta_{\perp}$ of $\le
0.1\deg$. Assuming this value, we would expect $\delta
T_{\parallel}\approx 3$ mK and $\delta T_{\perp}\approx 20\mu$K for the
D-band.

\subsection {Observed Offset}

\begin{figure*}
\plotone{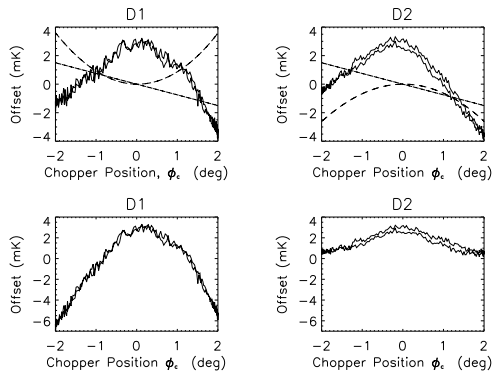}
\caption{
{\it Top Panels:} Measured chopper offsets, $O_i$, in D-band as
a function of chopper position for a typical night
of CMB data. This plot shows the raw data (with cal pulses subtracted)
synchronously coadded with the chopper azimuth, $\phi_c$.
Dashed lines are the calculated offsets due to
polarized emission from the chopping mirror with the offset at the
center of the chop removed, $T(\phi_c)-<T(\phi_c=0)>$. The 
straight lines
are the calculated offset due to a mirror tilt of
$\delta\theta_{\parallel}=0\deg.1$ in the plane
of the stationary chopper. When these offsets are multiplied by a
synthesis vector, the result is constant in time and independent of
chopper position. D1 is horizontally polarized; D2 is vertically
polarized. {\it Bottom Panels:}
The residual measured offset after the subtraction of the predicted
offsets. The form of the 
residual indicates that the source is polarized. The asymmetry in 
D1 may be due to feed rotation.} 
\label{fig:obs_offset}
\end{figure*}

Figure~\ref{fig:obs_offset} shows the chopper-position-dependent
offsets observed in each of the
D-band channels on a typical night analyzed for CMB observations in the
1998 campaign. Plotted in the top frames are the observed offsets (in antenna
temperature) as a function of azimuthal chopper angle.
Overplotted are the computed polarized emission offsets for each of the
channels (dashed line) and offset due to
mirror tilt about an axis parallel to the chopper normal (solid
line). The bottom frames show the offsets
corrected for these two effects. 

There is a clear asymmetry in the offset about $\phi_c=0$
in the top panels of Figure ~\ref{fig:obs_offset}. This is most likely
caused by a misalignment 
of the chopper with $\delta\theta_{\parallel}= 0.1\deg$ 
(equation~\ref{eq:deltat_parallel}), corresponding to the D-band beam
centroid moving up and down vertically
$\delta\Psi_{\parallel}=0\fdg 0053$ as it scans the
azimuth. This angle is just below the detection threshold of our measurements 
(Section~\ref{sec-obs}). 
With a zenith temperature of 10~K in D-band, 
$\delta\theta_{\parallel}= 0.1\deg$ 
produces a 1.6~mK modulation. 

Another mechanism for producing an asymmetry of the same magnitude
is the rotation of the polarization direction of a feed. A 
$\approx 1\deg$ rotation results in a signal of 3 mK in D-band 
(Table \ref{tab:offsets_summary}).
We cannot rule out that some part of the asymmetry is due to this,
though it would be coincidental to have the asymmetry so
similar in both feeds. In the HEMT channels, the signature of
the asymmetry due to feed rotation is opposite in the 
two polarizations. As the chopper sweeps, the emission for one
polarization goes up while that from the other polarization goes down. 
We see no evidence for such a signal.

The offsets for Ka and Q band are similar to those shown in
Fig.~\ref{fig:obs_offset} and in Fig. 1 of Herbig {\it et al.} 
(1998). Before accounting for the chopper emission, the 
magnitude is between 4 to 8 mK.
The offsets in Ka and Q bands in the Saskatoon experiment
were 1/2 to 2 mK, considerably smaller than those observed here.
We attribute the difference to the fact that the TOCO/QMAP 
system is, by necessity, much more compact: the chopper 
is closer to the feed horns
and the ground screens are closer to the main beam.

After accounting for the polarized chopper emission and the alignment
of the chopper, both of which can be computed accurately, 
there is still a residual polarized offset in all
channels of magnitude 2 to 8 mK.
In particular, the observed residual offset is always largest
in the horizontal polarization as shown for D-band in the 
bottom panels of Fig.~\ref{fig:obs_offset}. The offsets are largest 
in Q-band, and thus no particular characteristic spectrum is identified.
This suggests that the
source is modulated cavity emission, thermal emission from the
enclosure, or a combination of both.
We have not been able
to definitively identify the mechanism responsible for the effect
though the most likely source is the shallow angle diffraction
of the chopper cavity emission into the parabola.  Other potential
sources are either too small or do not have the correct polarization.

The offsets we discuss in this section are not atypical for
CMB experiments. They correspond to the raw detector output before any
of the symmetries or modulations of the experimental design have 
been utilized. In Sections 11 and 12 we discuss the offsets {\it after}
the strong spatio-temporal filter of the experimental method has 
been applied.

\section {Electronics and Data Acquisition}

\begin{figure*}
\epsscale{0.75}
\plotone{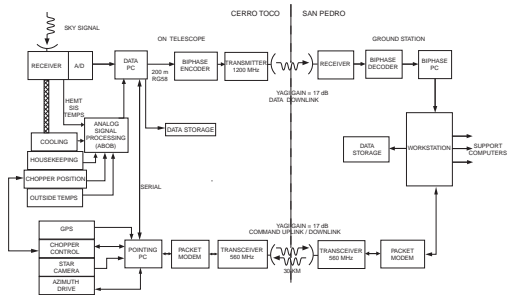}
\caption{Layout of electronics and telemetry system for TOCO.
The components on the left are on Cerro Toco and the ones on the right
are at the ground station in San Pedro (a hotel room in
La Casa de Don Tom\'as), at a comfortable living
altitude. The command up/down link is a 40 W Motorolla GM300 radio
communicating via a Paccomm packet modem.}
\label{fig:data_system}
\end{figure*}

Because it was balloon borne, QMAP was by necessity self contained.
It had its own command telemetry and CCD-based pointing system
\cite{dev98} though used the National Scientific Balloon Facility
(NSBF) transmitters to relay data. To operate remotely in Chile as the
TOCO experiment, a transmitter was added to telemeter a compressed
version of the data from the telescope on Cerro Toco 
to the ground station in  San Pedro de Atacama, 35 km away. 
A block diagram of the telescope electronics, data system, and telemetry
is shown in Figure~\ref{fig:data_system}.

\subsection{Electronics}

The eight radiometry signals are processed in the backpack.
Each signal is square-law detected with
detector diodes, amplified, and sent with differential line drivers
to a processing board. On the processing board, the signal is
high-pass filtered with a 2-pole RC network 
with $f_{3dB}=1~$Hz to remove 
the DC level which is separately recorded. This filter introduces
a small but stable electronic offset. The high-passed analog
signal is then digitized using an 18-bit
$\Sigma\Delta$ analog to digital converter
\footnote{One half of an Analog Devices AD1878}. The serial digital signal is
sent through a shift register and latch that converts it to a 32-bit 
parallel word accessible by a computer.
The use of $\Sigma\Delta$ ADCs is advantageous over
sampling ADCs because of their superior differential nonlinearity 
specification, which is
important when signals comparable to or smaller than the 1-bit level are of
interest. Additionally, these ADCs incorporate a digital anti-aliasing filter,
eliminating any temperature dependence or drift in this
component of the electronic band pass.

In the CMB analysis, we use HEMT data from 14 to 55 Hz ( $4<l<200$ ) 
and SIS data from 15 to 60 Hz ( $4<l<500$ ). 
The electronic bandpasses are defined on the 
lower end by the high-pass filter 
and on the high end by the $\Sigma\Delta$ chip. Over our frequency
range, the phase response of the chip is linear with frequency
and so the $\Sigma\Delta$ introduces a time delay in the signal. 
The amplitude response over the CMB frequencies varies by less 
than $0.4\%$ over this range.

\subsection{Computers and Telemetry}

Two single board computers, which handle the data and the
pointing, are located on the telescope. They are 
synchronized with a common clock and communicate remotely with the
computers operating at the base station. In the following, we focus on
the configuration for TOCO.

The ``Data PC'' logs the detector output and the position of
the chopping mirror as well as various voltages, currents, temperatures,
etc. which we use to monitor the telescope.
Two versions of the data are recorded. A complete version
($\approx 1$~G byte/day), which is used in the final analysis, is stored on
a 4~G bytes hard drive in
the data computer. This drive is contained in a pressurized vessel to
prevent damage resulting from operation at high altitude
\footnote{When not enclosed 
in a pressurized and dust-free container, most hard
drives, especially high capacity ones, were found to fail on a timescale
of a day.}. These drives
are swapped out every two to four days. The data are then uploaded 
onto the computer
system in the ground station, and stored on Exabyte tapes. A second
compressed version of the radiometry and housekeeping data is bi-phase encoded 
and sent real time to the ground station.

The ``Pointing PC'' records the position of the 17-bit absolute digital
encoder on the telescope azimuthal bearing, controls and
monitors the azimuthal drive motor, records the time from a GPS receiver,
and interprets commands sent remotely from the ground station.
A CCD camera and Matrox digital image processing board can track
stars during calibration and pointing verification procedures. 
The command status and all other information on the pointing
PC is passed to the data PC for logging. The Pointing PC can also reboot
the Data PC, an operation we sometimes found necessary. 

Two radio links allow us to communicate with the telescope from the
ground station. A high frequency link at $1.4$ GHz
(bandwidth of 100 kHz) with a
2~W transmitter links the data computer with a ground station
PC providing the bi-phase data. A marine radio operating at 460.5 MHz
communicates via a packet modem with the
pointing computer, providing commanding.

The ground station computers receive and
store the data, archive the data to tape, and run the commanding,
display, and alarm software. From the ground station, the telescope
can be slewed in azimuth, chopper parameters can be adjusted, 
and the cooling power (to
stabilize the temperature of the warm electronics) can be increased or
decreased. Most major systems can be turned on and off remotely. 

\section{Observations}
\label{sec-obs}

While both QMAP and TOCO were designed to measure the anisotropy,
their approaches were completely different. QMAP was designed to
make a true map of the sky. The data from the time stream were 
pixelized on the sky and the analysis was done on the sky map.
In the first flight of QMAP, the chopper swept horizontally 
at 4.7 Hz and the gondola
wobbled in azimuth with a period of 100~s about a meridian containing
the North Celestial Pole. This gondola motion, combined
with the rotation of the Earth, produced a highly interlocking 
scan pattern that allowed for the clean separation of instrumental 
effects from the celestial signal \cite{dOC98}.
 
\begin{table*}
\small{
\caption{Telescope and Chopper Motion}
\vbox{
\tabskip 1em plus 2em minus .5em
\halign to \hsize {
     #\hfil &#\hfil &#\hfil &#\hfil &#\hfil \cr
\noalign{\smallskip\hrule\smallskip\hrule\smallskip}
  Parameter& QMAP96a&QMAP96b& TOCO97& TOCO98\cr
\noalign{\smallskip\hrule\smallskip}
Chopper frequency, $f_c$ (Hz)&4.7&4.6&4.6&3.7\cr
Chopper amplitude in azimuth, $\phi_c$(max) (deg)&$\pm 10$&$\pm 2.5$&
        $\pm2.96$&$\pm 2.02$\cr
Elevation at center of chop, $\theta_{el}$ (deg)&40.7&40.1&40.5&40.63\cr
Azimuth of center of chop~\tablenotemark{a}, ~$\theta_{az}$ 
(deg) &scanned&scanned& 204.9&207.47\cr
Amplitude of wobble (s) & 100 & 50 & none & none \cr
Amplitude of wobble in azimuth (deg) & $\pm5^{\circ}$ & $\pm1\fdg5$
& none&none \cr
\noalign{\smallskip\hrule\vskip1pt\hrule\smallskip}
}}}
\tablenotetext{a}{This is the physical motion in a horizontal plane. The
amplitude on the sky is $2\phi_c\cos\theta_{el}$.}
\label{tab:observing_parameters1}
\end{table*}

\begin{figure*}
\epsscale{0.6}
\plotone{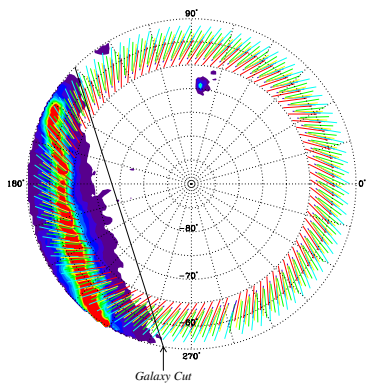}
\caption{Scan pattern for the TOCO98 campaign around the SCP. 
The scans are shown as straight lines for
each feed for 15 minute intervals around the sky.
A more detailed picture would show the lines with slight curvature
to account for the projection.
The center of the chop is at an azimuth of $207\deg.5$ and an elevation
of $40\deg.6$. The chopper sweeps out $6\fdg12$ on the sky
as it scans in azimuth. 
This results in a ring around the SCP approximately centered on 
$\delta=-61\deg$. The map of the galaxy is from 
Schlegel {\it et al.} (1997). }
\label{fig:skyscan}
\end{figure*}

The TOCO scan was designed to measure the angular power spectrum\footnote{We
decided against observing on both sides of the South Celestial Pole
(which would have produced interlocking scans thereby facilitated 
map production) to maximize the stability of the instrument and to 
minimize the complexities of the analysis.}. The telescope
optical axis is fixed in azimuth and elevation, as indicated in 
Tables~\ref{tab:centroids} \&
\ref{tab:observing_parameters2}, and the chopper sweeps the beam
across the sky. The beams cover an annulus around the SCP as shown in 
Figure \ref{fig:skyscan}.

The TOCO observing site is located at an altitude of 5200 m on
Cerro Toco in the Northern Atacama desert in Chile near the
borders of Argentina and Bolivia. The latitude is
$22\deg.95$ South and the longitude is $67\deg.775$ West. 
A building for an abandoned sulpher mine blocks the occasional
60 kt winds. The Atacama is one of
the highest, driest deserts in the world and is therefore a good place
for millimeter and centimeter wave observations.
We find that the weather is good enough for D-band CMB
observations $\approx 50\%$ of calendar time between September
and January.

\begin{table*}
\small{
\caption{Observing Parameters}
\vbox{
\tabskip 1em plus 2em minus .5em
\halign to \hsize {
     #\hfil &#\hfil &#\hfil &#\hfil &#\hfil &#\hfil &#\hfil &#\hfil &#\hfil \cr
\noalign{\smallskip\hrule\smallskip\hrule\smallskip}
  Campaign& Ka1&Ka2&Q1&Q2&Q3&Q4&D1&D2\cr
\noalign{\smallskip\hrule\smallskip}
 QMAP96a  \cr
Sample frequency, $f_s$ (Hz)&752&752&752&752&752&752&...&...\cr
Samples per chop, $N_s$&160&160&160&160&160&160&...&...\cr
$\Theta_{\rm FWHM}$ per scan&20&20&30&30&30&30&...&...\cr
Samples per $\Theta_{\rm FWHM}$
      (F/S)\tablenotemark{a,b}&0.78/3.2 &.78/3.2 &0.56/2.7 &0.56/2.7 
                              &0.58/2.7 &0.58/2.7 &...&...\cr 
Point Source Sensitivity $\Gamma(\nu_c)(\mu {\rm K Jy}^{-1})$&128&128&
   132&132&129&129&...&...  \cr
\noalign{\smallskip}
 QMAP96b  \cr
Sample frequency, $f_s$ (Hz)&736&736&736&736&736&736&...&...\cr
Samples per chop, $N_s$&160&160&160&160&160&160&...&...\cr
$\Theta_{\rm FWHM}$ per scan&5&5&8&8&8&8&...&...\cr
Samples per $\Theta_{\rm FWHM}$ (F/S) &3.1/6.5 &3.1/6.5 &2.4/5.7
                                      &2.4/5.7 &2.3/5.6 & 2.3/5.6&...&...\cr
Point Source Sensitivity $\Gamma(\nu_c)(\mu {\rm K Jy}^{-1})$&136&136&
   146&146&120&120&...&...\cr
\noalign{\smallskip}
 TOCO97$^a$   \cr
Sample frequency, $f_s$ (Hz)&368&368&368&368&368&368&1472&...\cr
Sky detection frequency (Hz)&18-55&18-55&18-64&...&23-64&23-64&...&... \cr
Samples per chop, $N_s$&80&80&80&80&80&80&320&...\cr
$\Theta_{\rm FWHM}$ per scan&6&6&9&9&9&9&30&...\cr
Samples per $\Theta_{\rm FWHM}$\tablenotemark{b}
(F/S)&1.3/3.0& 1.3/3.0 &1.0/2.6 &1.0/2.6 &1.0/2.6 &1.0/2.6&1.1/5.2&1.7/6.7\cr
Point Source Sensitivity $\Gamma(\nu_c)(\mu {\rm K Jy}^{-1})$&131&131&
   128&120&119&118&87.3&49.5\cr
\noalign{\smallskip}
 TOCO98$^a$   \cr
Sample frequency, $f_s$ (Hz)&296&296&296&296&296&296&1184&1184\cr
Sky detection frequency (Hz)&...&...&...&...&...&...&19-59&19-63 \cr
Samples per chop, $N_s$&80&80&80&80&80&80&320&320\cr
$\Theta_{\rm FWHM}$ per scan&4&4&6&6&6&6&20&20\cr
Samples per $\Theta_{\rm FWHM}$ (F/S)& 1.9/3.6 &1.9/3.6 &1.5/3.2 &1.5/3.2
                                     & 1.4/3.0 &1.4/3.0 &1.7/6.7 &2.5/8.1\cr
Primary Aperture Efficiency, $\eta_p$&
        0.58&0.58&0.65&0.62&0.56&0.55&0.52&0.24\cr
Chopper Aperture Efficiency, $\eta_p$&
        0.10&0.10&0.11&0.11&0.10&0.10&0.087&0.040 \cr
Point Source Sensitivity $\Gamma(\nu_c)(\mu {\rm K Jy}^{-1})$&121&121&
   138&131&119&116&117&54.7\cr
\noalign{\smallskip\hrule\vskip1pt\hrule\smallskip}
}}}
\tablenotetext{a}{F/S refer to the fast and slow parts of the sinusoidal
motion near the center and edges of the chop.}
\tablenotetext{b}{The sky is not Nyquist sampled at the center of the
chop. This is accounted for in the mapmaking and beam synthesis. 
The $e^{-1}$ point of the beams, $l_e=\sqrt{16\ln2}/\theta_{FWHM}$, is
$l_e=212, 273,~{\rm and}~955$ for Ka, Q, and D respectively. The
undersampling for QMAP96a is severe and limits the 
map reconstruction in the current pipeline to $l\approx 180$.}
\label{tab:observing_parameters2}
\end{table*}

\section{Calibration}
\label{sec-calibration}

Calibration uncertainty is the largest source of experimental error for 
both QMAP and TOCO as it is for many other CMB experiments.
In general, calibration is dominated by systematic effects.
Multiple independent measurements are required to determine the
uncertainty. Short of this, one's knowledge of the
experiment is called upon to set the uncertainty and its distribution
\footnote{One may, for example,
conservatively use the difference between two measurements as the
uncertainty (e.g., de Bernardis {\it et al.} 2000) even though 
the formal uncertainty is 
the difference divided by $\sqrt{2}$. To determine error bars 
on quantities derived from the data, one marginalizes over the 
distribution that describes
the calibration. Generally the distribution is taken as Gaussian,
though one might reach different conclusions if a different
distribution in fact described the data.}.

A calibration for TOCO involves the following steps: a) position
the telescope so that Jupiter either rises or sets through the azimuthally 
swept beam; b) make a map of the source and find the best fit 
amplitude, position, and beam profile; c) compute the brightness of
Jupiter accounting for its position with respect to Earth; 
d) correct for slow drifts in calibration, correct for the 
electronic response of the receivers, and convert Rayleigh-Jeans to
thermodynamic temperature.
A similar procedure was used for QMAP. In the following, we consider
each element of the calibration and its associated uncertainty.

There are two general classes of calibrators, those that fill the beam and
point sources. FIRS \cite{Ganga93}, BOOMERanG, and
MAXIMA used the dipole as calibrated by 
both FIRAS and DMR aboard {\sl COBE}. The dipole signal corresponds to angular
scales larger than, and detection frequencies lower than, those used for
CMB data analysis. Thus knowledge of the electronic transfer
function, beam profile, and spatial filtering are necessary for the
extrapolation. At the other extreme, point
sources are at smaller angular scales and higher post-detection frequencies
than those used for CMB analyses. They have the advantage that the calibration and
beam profile are measured simultaneously. Still, though, one must
account for the electronic transfer function and spatial filtering. 

QMAP was calibrated with Cas-A\footnote{3C461, 2321+583IAU(1950), 
$l=111\fdg7$, $b=-2\fdg1$} and TOCO was calibrated using Jupiter.
As the beam sweeps across the source, the detector output is given by
\begin{eqnarray}
\mathcal{C}V(t)={1\over 2}\int_{\nu}\int_{\Omega}\int_{t^\prime} A_e(\nu)g(\nu)
S_\nu(\vec\Omega)P_n(\nu,\vec x-\vec x^\prime(t^\prime))\times
\nonumber\\~~~~I(t-t^\prime)d\vec x d\nu dt^\prime~~~
\label{eq:cal1}
\end{eqnarray}
where $V$ is the voltage out of the detector, $A_e$ is the effective 
area of the telescope dish, $g(\nu)$ is
the passband of the receiver, $S_\nu$ is the source
surface brightness in units of ${\rm W/m^2\,sr\,Hz}$, $P_n$ is the 
normalized gain of the antenna, $\vec x$ is the direction on the sky,
and $I$ is the impulse response function
of the electronics\footnote{A typical frequency dependent loss 
in the system has a negligible affect on the calibration.}.
Here $\mathcal{C}$ is the calibration constant that relates the
source temperature to the measured voltage. We do not explicitly account
for sampling in the integral.  

Each term in equation \ref{eq:cal1} has an error associated with it that
contributes to the net uncertainty. The effects include:
a) $\sigma_S$, the intrinsic uncertainty in the brightness 
temperature of the source
at the calibration frequency extrapolated from previous
measurements; b) $\sigma_V$, the
uncertainty in the measured temperature of the source; 
c) $\sigma_\Omega$, the
uncertainty in the measured solid angle of the beam
which includes any beam smearing; d)the
uncertainty in the measured receiver bandpass, $\sigma_g$, and 
center frequency, $\sigma_{\nu}$; e) $\sigma_{drift}$,  the uncertainty in
the change of the calibration between when it is measured and when it is
applied; and f) $\sigma_I$, the uncertainty in
the frequency response of the instrument. 
Items (b) and (c) are derived
from measurements of the source and are covariant; generally one
conservatively treats them as independent errors.
Combining these sources, the intrinsic calibration is given by:
\begin{eqnarray}
\biggl({\sigma_c\over C}\biggr)^2 = 
\biggl({\sigma_S\over S}\biggr)^2 + 
\biggl({\sigma_V\over V}\biggr)^2 +
\biggl({\sigma_\Omega\over \Omega}\biggr)^2 + \nonumber\\
\biggl({\sigma_{drift}\over drift}\biggr)^2 + 
\biggl({\sigma_I\over I}\biggr)^2
\label{eq:cal2}
\end{eqnarray}
We evaluate these terms in the following and summarize the results
in Table~\ref{tab:cal_errors}.

\subsection{Calibration Source Brightness Temperatures}
\label{sec-int}

The flux from a source is typically given as $f_{\nu}=\int
S_{\nu}(\theta,\phi)d\Omega$ and is reported in units of Janskies
(${\rm 1~Jy = 10^{-26}~W/m^2\,Hz}$) as a function of frequency.
A power-law model is used to extrapolate the calibration frequencies 
to the observation frequency, $f_{\nu}=f_0(\nu/\nu_0))^{\beta}$. When the
source angular size is a significant
fraction of the beam or is changing in time, as it is for planets, it is more
convenient to use the brightness temperature.
 
\subsubsection{Cas-A Flux for QMAP}

Cas-A is unresolved at the QMAP beam size and the correction for its finite 
size is negligible. From a combination  
of the data \cite{Baars77,Chini94,Mezger86}, we find 
$\log(S_{\nu}/{\rm Jy})=(5.713 \pm 0.023) -(0.759\pm 0.006)\log(\nu/{\rm
MHz})$
at epoch 1980 \cite{Herbig98}. 
At 31.3 GHz, the mean frequency of the QMAP Ka bands, the flux corresponds to
199.9 Jy.  Baars {\it et al.} (1977) give the percentage annual decrease as
$\delta S_\nu/S_\nu =0.97(\pm 0.04)-0.3(\pm 0.04)\log(\nu/{\rm GHz})$
and so this value is reduced to 183~Jy for epoch 1997. When one takes 
into account
all of the above errors, including $\sigma_\nu$, the uncertainty is 
$(\sigma_S/S)=8.7$\% for both Ka and Q bands.

After the QMAP data release, Mason {\it et al.} (2000) 
reported $S_{cas,1998}=194\pm
4.7~$Jy at 32 GHz which we convert to 195 Jy for epoch 1997. This
measurement, which is very close to our frequency, greatly reduces the
uncertainty associated with the interpolation and secular decrease. When
the errors in the central frequency are included (Table~\ref{tab:centroids}), the 
uncertainty is $(\sigma_S/S)=2.7$\%. 
For Q band, the flux is $159\pm 4.8~$Jy. The
slightly larger error is from the extrapolation from 32 to 41 GHz.
The result is an increase in the temperature scale of the 
QMAP data by 6.6\% and reduction in the calibration uncertainty, $\sigma_S$.  

\subsubsection{Jupiter Temperature for TOCO}

The brightness temperature of Jupiter is measured
by Ulich (1981), Griffin {\it et al.} (1986), Ulich and
Mason {\it et al.} (2000), and is taken to be 152, 
160, and 170 K in Ka, Q, and D band respectively
with an intrinsic calibration error of $\sigma_S/S=5$\%. As the 
temperature is a
weak function of frequency across our bands, the uncertainties resulting from 
$\sigma_\nu$ and $\sigma_F$ are negligible. 
The Jupiter calibration temperature is
obtained by scaling the brightness temperature by the ratio of the
solid angle of Jupiter (determined from ephemerides) to the measured
solid angle of the beam. A typical measured temperature of Jupiter is 
15~mK, 30~mK, and 350~mK in Ka through D1 bands respectively.

Finally, one applies a correction to convert small changes in antenna 
temperature to small changes in thermodynamic CMB temperature. We use:
\begin{equation}
\eta=\bigg({\partial T_{\rm ant}\over\partial T_{CMB}}\bigg)
={x^2e^x \over(e^x-1)^2}
\label{eq:eta}
\end{equation}
where $x=h\nu/kT_{CMB}$.
For the TOCO98 D-band data, the data are
multiplied by $\eta^{-1}$, or 1.66 for D1 and D2 to convert 
from data calibrated on Jupiter's brightness
temperature to thermodynamic units referenced to the CMB. 
For the Ka and Q band data we multiply by 1.02 and 1.05 respectively.
The error in these measurements depends on the knowledge of the
band centroids and introduces
$\leq 1 \%$ uncertainty.  

\subsection{Measured Beam Solid Angle and Temperature}
\label{sec-ang}
To convert the measured fluxes to a temperature, the beam solid angle
must be known. Table~\ref{tab:beams} 
gives the beam determination for all campaigns along with the results 
of a computer model. For QMAP, the solid angle was determined from one 
in-flight mapping of Cas-A \cite{Herbig98}. The statistical error on the 
fit varied between 1\% and 3\% depending on the flight and band. 
As the beam fitting includes modeling of the instrumental offsets,
there is additional systematic error resulting in a net uncertainty
of $\sigma_\Omega/\Omega=5~$\%. For 
TOCO, the beam solid angles were determined
from a global fit to the Jupiter calibrations made during good
weather. This was done separately for both campaigns. In total, Jupiter
was mapped over 70 times. All results from the four campaigns are
consistent with our models. From all of our measurements of the beams, we 
conclude that the error on the solid angle for Ka and Q bands is 5\%,
for D1 it is 5\%, and for D2 it is 5.5\%. 
These values are dominated by systematic errors.

It is often convenient to parameterize the beams with a two dimensional
Gaussian profile. For Ka and Q bands, this introduces a negligible error.
We tested for this in TOCO97 \cite{Torbet99} analysis where we used the
measured Ka and Q-beam profiles in place of the parameterized
profiles and found $<1$\% difference in the final results.
The D-band channels are less well approximated by Gaussian profiles and thus
one must use the measured profile for accurate results as was done in 
Miller {\it et al.} (1999).

For each calibration, a seven parameter model is fit to the data.
We find the position of the source, the best 2-D Gaussian 
parameterization including orientation, the amplitude, and an offset. 
For QMAP, the 1-2\% statistical error on the amplitude is dominated
by a 3\% systematic uncertainty in the algorithm to extract the 
amplitude. In the TOCO experiment, the standard deviation of the fitted amplitude
is 4-10\% for all the HEMT channels in both seasons. The variance is
a result of atmospheric fluctuations and finite HEMT sensitivity and thus
averages down as the square 
root of $\approx 20$ independent high quality maps in each 
season to a value of 2\%.
There is a small additional uncertainty due to the fact that not all 
fitting algorithms give the same results. The net result is to increase 
the uncertainty in the amplitude to $\sigma_V/V=3$\%. 

\subsection{Calibration drift}
\label{sec-drift}

The physical temperature of the TOCO instrument can vary by 50~K in a day.
Even though all critical components are thermally regulated, there
are temperature changes that lead to changes in gain. 
Jupiter was observed, on average, within
two hours of the beginning of the CMB observing time. Changes in
system gain on time scales shorter than 24 hours were monitored with
an internal calibration signal (``cal pulse'')  with an effective 
temperature of $\approx 1$ K in all bands. This pulse was turned on 
40 msec twice every 200
seconds. The amplitude of the pulse was regressed with the body
temperatures
of the noise sources, warm electronics temperatures, and cryogenic 
temperatures in all bands.
The fit coefficients are consistent with laboratory
measurements and show that the pulse amplitude is constant
but that the system gain is a function of the microwave 
amplifier temperatures.

For TOCO, a typical long term (50 days) variation is 15\% in Ka band, 
5\% in Q band, and 20\% in D-band. The cal pulse amplitudes 
follow the general
trends in the Jupiter calibrations in all bands. From the cal pulses and
the Jupiter observations, we
derive a calibration drift model which we apply to the data.
In D-band, atmospheric fluctuations made 
use of the cal pulses to correct drifts over periods of less than six
hours problematic. The uncertainty in the model is estimated to be 5\%.  

A similar approach was taken with QMAP though the cal pulses were
clearer and so the uncertainty in correcting for the 5\% drift is 
negligible.
 
\subsection{Electronic passband}

Observing a point source is similar to exciting the electronics 
with a pulse. Consequently there are frequency components up to 
$f\approx 4[\phi_c\cos(\theta_{el})\theta_{FWHM}]f_c$
(Table~\ref{tab:observing_parameters1}) of the post-detection 
electronics. The CMB signal
is at comparatively lower frequencies. We model the full electronic
response of the system, including sampling, and find that the CMB
data should be reduced by 1.7\% in D-band, 1.5\% in Q-band, and 1\%
in Ka-band for TOCO and 1.5\% for QMAP over what one would get without the
correction. We estimate the uncertainty in this to be 1\% for 
TOCO and 5\% for QMAP. These shifts were not reported in the original 
papers as they were much smaller than the uncertainty, though we
include them here.

From Table~\ref{tab:observing_parameters2}, one sees that we are
slightly undersampled during the fastest part of the chop (where the
least amount of time is spent). To check for a possible systematic
effect in TOCO associated with this, calibrations were done with the source 
at the center of the chop and off to one side and with different chopper
amplitudes. The results of these tests are statistically consistent
with the nominal calibrations. For TOCO, the CMB anisotropy results are
insensitive to the slight undersampling because the CMB detection
frequency is far below the
sample frequency--due to the beam synthesis--and because in the beam
synthesis we simulate the sky scan. 

For the first flight of QMAP the 
undersampling is more drastic and was not included in detail in
the mapmaking. Thus, the QMAP data should only be considered valid
up to $l=180$\footnote{Neither the undersampling nor the calibration
shift were accounted for in Xu {\it et al.} (2001) and 
Wang {\it et al.} (2001).}. 
For both flights,
the calibration data were processed in a manner similar
to the mapmaking and so beam smearing effects were
accounted for in an average sense.

The phase response of the full instrument as a function of 
frequency was measured in the lab
and determined from observations of the Galaxy and Jupiter.
We find that the phase is linear over the range of
frequencies applicable to both CMB and point source observations.

\subsection{Combining the calibrations between bands and systematic
shifts from previous results.}

Both QMAP and TOCO have multiple detectors, the measurements of which are
combined into one angular spectrum. The net calibration uncertainty
is a combination of terms that are completely correlated between
channels, such as $\sigma_S$, $\sigma_\Omega$, $\sigma_{drift}$, \& 
$\sigma_I$, and terms that 
are uncorrelated, such as $\sigma_V$ and $\sigma_\eta$. When the data 
are combined the last two terms become negligible. The uncertainties
are 8\% for D1+D2, for the TOCO HEMTs, and for the combination of the
D1+D2+HEMTs. For QMAP the net uncertainty is 7.6\%. In the regions where
the QMAP and TOCO angular spectra overlap, only $\sigma_S$ and 
$\sigma_\Omega$ are correlated and the combined uncertainty is 6.4\%.

These uncertainties are slightly different than those previously
quoted for these experiments and are the result of a complete
reassessment of the calibration errors. To correct the previously 
published results, the
QMAP data should be multiplied by 1.051, The TOCO D-band
data \cite{Miller99} should be multiplied by 0.983, and the TOCO HEMT
data \cite{Torbet99} by 0.99.

\begin{table*}[bt]
\caption{Summary of Contributions to Calibration Uncertainty for
Individual Channels}
\small{
\vbox{
\tabskip 1em plus 2em minus .5em
\halign to \hsize {
     #\hfil &#\hfil &#\hfil &#\hfil &#\hfil &#\hfil &#\hfil  &#\hfil \cr
\noalign{\smallskip\hrule\smallskip\hrule\smallskip}
{\rm Channel}&{$\sigma_S~(\%)$}&{$\sigma_\Omega~(\%)$}&
{$\sigma_V~(\%)$}&{$\sigma_{drift}~(\%)$}&{$\sigma_{\eta}~(\%)$}
&{$\sigma_I~(\%)$}& {$\sigma_\mathcal{C}$~(\% )} \cr
\noalign{\smallskip\hrule\smallskip}
D1 & 5& 5.5 &3 &4 &1 &1 & 9.1 \cr
D2 & 5& 5 &3&4 &1 &1 & 8.8 \cr
TOCO HEMTs &5 &5 &3 &4 &$\cdots$ &1 & 8.7\cr
QMAP HEMTs &2.7 &5 &3 &$\cdots$ &$\cdots$ &5 &8.1 \cr
\noalign{\smallskip\hrule\vskip1pt\hrule\smallskip}
}}}
\tablenotetext{}{The intrinsic Jupiter
calibration uncertainty is given by $\sigma_S$
(Section~\ref{sec-int}), $\sigma_\Omega$ is the
uncertainty in measured solid angle (Section~\ref{sec-ang}), $\sigma_V$ 
is the uncertainty in measured Jupiter brightness temperature 
(Section~\ref{sec-ang}), and $\sigma_{drift}$
(section~\ref{sec-drift}) is the
uncertainty due to the calibration drift between 
observations, $\sigma_{\eta}$ refers to the Rayleigh-Jeans to
thermodynamic conversion due to the uncertainty in the centroid,
and $\sigma_I$ is the uncertainty on the electronic bandpass correction.
These calibration errors are in temperature.} 
\label{tab:cal_errors}
\end{table*}

\section{Observing the Anisotropy}

The anisotropy is a two dimensional random field in temperature.
The goal of CMB anisotropy experiments is to measure the characteristics
of that field. The three methods in use are direct mapping, time-domain 
beam synthesis, and interferometry. For a small number of detectors,
direct mapping makes the most efficient use
of them, beam synthesis is the next most efficient, and
interferometry is the least efficient, as we discuss in 
Section~\ref{sec-demod}. The best strategy to use, though,
depends as much on the control of potential systematic errors as
on raw sensitivity. 

The QMAP experiment was designed to make a direct map. By this we mean
that the time ordered data are assigned a sky pixel number as they come out of
the detector. Slow drifts in the detector output may be removed from the
map using a variety of methods \cite{Cottingham,dOC98,Hivon01}. For all
methods, though, a heavily interlocking scan strategy is required for 
robust results. QMAP produced multiple  $\approx 0\fdg8$ resolution
maps of the CMB from two flights. These maps were found to be 
consistent with each other and with the Saskatoon maps \cite{Xu00}.
In other words, two entirely separate experiments measured the same
temperature variations in the sky in overlapping regions. The other
mapping experiments to do this are COBE/DMR \cite{smoot92} with
FIRS \cite{Ganga93}, and  COBE/FIRAS with COBE/DMR \cite{Fixsen97}.
A wide range of systematic checks have been applied to QMAP
and it passes them all.

The TOCO experiment used time-domain beam synthesis\footnote{Atmospheric
fluctuations preclude direct mapping from the ground.}. 
To our knowledge, the method was first employed in Netterfield {\it et
al.} (1995). Since the Saskatoon experiment, we have refined 
the technique, applied it to a multifeed system, and incorporated 
numerous cross checks and systematic checks of the robustness 
of the solution. Most of the remainder of this paper is devoted 
to describing those checks.

\section {Time Domain Beam Synthesis}
\label{sec-demod}

As with all ground-based CMB experiments, 
the effects of atmospheric fluctuations must be strongly suppressed. TOCO
uses the chopping flat to scan the beam across the sky in a
sinusoidal pattern (Section~\ref{mirrors_groundscreens}).
In a postdetection analysis, the time ordered data are multiplied
by synthesis vectors, $SV_{n,i}$, that have half the period of the chopper
cycle. Thus, the data with the chopper moving in one direction
is coadded with the data with the chopper moving in the other direction.
We form
\begin{equation}
t_n=\sum_{i=1}^{\rm N_c}SV_{n,i}d_i
\label{eq:demod}
\end{equation}
where $d_i$ is the vector containing the raw data from a full 
chop cycle,
$n$ is the index or ``harmonic'' of the synthesized beam,
and $N_{c}$ is the number of samples in a chop cycle.

There is no set prescription for $SV_{n,i}$. The best choice
depends on the scan pattern (e.g., sinusoidal or triangular),
the desired degree of orthogonality between synthesized beams,
the shape of the resulting window function, and the orthogonality
to any potential offset. For instance, one may pick  $SV_{n,i}$
so that $t_n$ is insensitive to the secant dependence of the 
atmospheric gradient. For the 
sinusoidal scan patterns, we find a useful set is given by
\begin{equation} 
SV_{n,i}\propto \cos(\pi (n-1)(1-\sin(2\pi(i-1/2)/N))/2) ,
\nonumber\\~~~~i=1,...N_c~~~~~~~~~~~~
\label{eq:wv}
\end{equation} 
though ultimately we tune the $SV_{n,i}$.
The synthesis vectors are effectively apodized sine functions
each of which is designed to produce a different effective antenna
pattern, and thus probe a different angular scale. 

The phase of the electronic signal with respect to 
the position of the beam on the sky is determined by forming the quadrature signal
(data from the first half of chopper sweep minus data from the second
half resulting in minimal sensitivity to celestial signals)
as a function of phase for each harmonic. Then the phase
that nulls the galactic signal over twenty-five of the best 
observing days is found. 
There is a small harmonic dependent component to the best fit
phase that is well modeled with a linear fit.
The phase shift is incorporated into equation~\ref{eq:wv}
for all analyses. The galaxy-null determined
phases agree with the phases determined from Jupiter observations, derived
with completely independent code. In addition, the entire 
data analysis is redone after setting the phase ahead and behind the best fit 
by twice the error derived in the fit.
No changes in the final results are seen.

The resulting effective antenna
sensitivity patterns, or synthesized beams $H({\bf x})$, are given by
\begin{equation}
H({\bf x})=\bigg<\sum_{i} SV_iG({\bf x}-{\bf X_i})\bigg>_{\rm RA~bin}
\label{eq:antenna_sensitivity}
\end{equation}
where the center of the main beam is located on the sky at position
${\bf X_i}$, $SV_i$ is the synthesis vector, and 
\begin{equation}
G(x,y)={1\over 2\pi\sigma _x\sigma _y}{\rm exp}\bigg(-{x^2\over
2\sigma _x^2}-{y^2\over 2\sigma _y^2}\bigg)
\label{eq:G}
\end{equation}
is the main beam pattern of the telescope pointed at the center of the
chop. Here $x$ measures the position in the
azimuthal direction, $y$ changes with elevation, and the beamwidth of the
telescope is $\sigma =\theta_{\rm FWHM}/ \sqrt{8{\rm ln}2}$.
Beam smearing due to the finite size of the sky bins
is incorporated into equation~\ref{eq:antenna_sensitivity}. As tests, 
the A/D sampling and the change in the effective horizontal beam width
as a function of chopper speed are included in equation~\ref{eq:G}. These effects
are negligible. As noted earlier, equation~\ref{eq:G} was replaced
with the true beam profile for D band. This was not necessary for the other bands.
The synthesized beams are then normalized, by adjusting the amplitude
of $SV_{n,i}$, so that:
\begin{equation}
\int{|H({\bf x})|d{\bf x}}=2
\label{eq:beam_normalization}
\end{equation}
In summary, the synthesis vectors and beam synthesis incorporate all
known aspects of the motion of the beam on the sky.

Data are binned according to right ascension at the center of the
chopper sweep into 768 ``fundamental bins'' around the complete circle
shown in Figure~\ref{fig:skyscan}. The fundamental bins are subgrouped
into $N_{RA}$ right ascension bins where $N_{RA}$ depends on the
harmonic. Thus, the average in equation~\ref{eq:antenna_sensitivity}
depends on harmonic. To avoid statistical
bias, it is important that the mean, sample variance, and error on the mean
in each of the $N_{RA}$ bins is computed for {\it all} the data that land in
a bin in a given night. For example, for the 1998 4-pt\footnote{Following the
notation in Netterfield {\it et al.} (1997), 
we call a synthesized beam with n lobes an
n-pt beam (see Figure~\ref{fig:1focal_plane}).}, 
Q-band beam, there are $N_{\rm RA}=20$ bins around the circle and so
in one night, $\approx 1.7\times10^{4}$ values of $t_4$ 
(equation~\ref{eq:demod}) are averaged
together for each bin. At low $l$, the variance in any RA bin is larger
than that expected from detector noise alone due to atmospheric
fluctuations. For integrations longer than roughly 3 minutes, 
the noise is stationary for the cuts described below. 
Finally, the results from individual nights are averaged.

The harmonics are analyzed individually at low $l$ or in groups at
higher $l$. Data from different detectors are also combined. 
The full theory covariance matrix, $C_T$, is computed for each harmonic or 
combination \cite{Bond96} along with the Knox filter \cite{Knox99}
and effective spherical harmonic index, $l_e$, of 
the observations.
The noise covariance matrix, $C_N$, is computed from the data. Other
than detector noise, which is uncorrelated between bands, the dominant
contribution is the atmosphere. Unlike Saskatoon, the frequency bands
are not subdivided and there is no East-West chopping.

For each group of harmonics, we find the likelihood as a function of the
band power, $\delta T_l^2$, according to\footnote{In this paper, we
report the band power following Bond as opposed to Netterfield.
The difference is $2(l+1)/(2l+1)\approx 0.2~$\% at $l=200$}:
\begin{equation}
L(\delta T_l^2)={1.\over 2\pi^{N/2}|\mathbf{M}|^{1/2}}
\exp(-t^T\mathbf{M}^{-1}t/2),
\label{eq:like}
\end{equation}
where $\mathbf{M}=C_T(\delta T_l^2)+C_N$. All the analysis is
done as a function of $\delta T_l^2,$ though we report
$\delta T_l$ because it gives a direct measure of signal-to-noise
as the detector output is proportional to temperature.

It is sometimes convenient to estimate the signal-to-noise
for a single measurement for a single harmonic. This 
can then be generalized to a set of N independent 
measurements \cite{Knox95}.
The measured {\it rms} amplitude of the sky fluctuations is given by
$\Delta = \delta T_l\sqrt{I(W)}$ where $I(W)=\sum_l(W_l/l)$, with 
$W_l$ the window function, encodes the efficacy of the synthesized 
beam. The noise of any measurement is given by
$\kappa_n\tilde T/\tau_{obs}^{1/2}$ where
\footnote{For the classic single difference, $\kappa=\sqrt{2[1^2+(-1)^2]}=2$.}
$\kappa_n=\bigg[N_c\sum_{i=1}^{N_c}SV_{n,i}^2\bigg]^{1/2}$,
$\tau_{obs}$ is the time spent observing a point, and
$\tilde T$ is given by equation~\ref{eq:det_noise}. 
From fits to the computed $I(W)$, we find
$\sqrt{I(W)}\approx 2.6/n^{0.75}\approx 6/\sigma_b^{0.25}l_e^{0.75}$
where $n$ denotes the $n$-pt function, $\sigma_b$ is the Gaussian width
of the beam, and $l_e$ is the associated effective
spherical harmonic index. The signal-to-noise for a particular
measurement is then
\begin{equation}
S/N = {\delta T_l\sqrt{I(W)}\tau_{obs}^{1/2}\over \kappa_n \tilde T} 
\approx  
{0.7\over \sigma_b^{0.5}l_e^{0.8}}
{\delta T_l\tau_{obs}^{1/2}\over \tilde T} 
\label{eq:tdbs_sn}
\end{equation}
We may interpret this result as saying that for a fixed sensitivity, 
$\tilde T$, and a flat angular spectrum, $\delta T_l$, beam synthesis
reduces the effective temperature of the sky by
$\gamma^{TDBS}=0.7/\sigma_b^{0.5}l_e^{0.8}$. This form fits Ka-band through D-band 
data to 30\% accuracy for TOCO
\footnote{Other functional forms work as well.
This is presented to aid in estimating the S/N. For SK,
which used different criteria to synthesize the beams,
$\gamma^{TDBS}=3.6/ \sigma_b^{0.5}l_e$, roughly a factor of two higher
at low $l$. However, the window functions were less well 
localized.}. 

It is worth contrasting beam synthesis in interferometry with
time domain beam synthesis. In interferometry, each baseline
yields one synthesized beam enveloped by the primary beam pattern
of a single element.
A similar expression to equation~\ref{eq:tdbs_sn} obtains for 
interferometers with Gaussian main
beams of width $\sigma_b^{Int}$ \cite{Hobson95,White98}.
We find 
\begin{equation}
S/N = {\sqrt{2}\over \sigma_b^{Int} l_0}
{\delta T_l\tau_{obs}^{1/2}\over \tilde T} 
=\gamma^{Int} {\delta T_l\tau_{obs}^{1/2}\over \tilde T}
\label{eq:int_sn}
\end{equation}
where $l_0 = 2\pi u_0$ is determined by the separation of the 
two antennae. For the 5-pt Q-band synthesized beam, 
$\theta_{FWHM} = 0\fdg7$ ( $\sigma_b = 0.0052$ )
and $l_e=87$. Similar coverage in $l$-space would be obtained
with an interferometer with $\theta_{FWHM}=4\fdg86$ 
( $\sigma_b^{Int}  = 0.036$) and $l_0=87$. 
From these, $\gamma^{TDBS} = 0.27$ and 
$\gamma^{Int}= 0.45$. To the level of accuracy of the fitting 
functions, these are equivalent. 

The sensitivity advantage of time domain beam
synthesis is that multiple n-pt functions are measured 
simultaneously with a single detector. 
In the parlance of Fourier transform spectroscopy, there is a multiplex
or ``Felgate'' advantage over an interferometer with just a 
few antennae. However, as the number of interferometer 
baselines scales as $n_a(n_a-1)/2$, where $n_a$ is the number of
antennae, large interferometers achieve high sensitivity
\cite{pad00,Pryke01}.
For the ideal mapping experiment, with minimal baseline subtraction,
the advantage over both interferometry and time domain beam synthesis 
is that in a fixed amount of time more spatial modes can be measured.

\section{Data Selection}
\label{data_selection}

Most of the analysis effort goes into data selection and 
testing to make sure that the selection does not bias the final result.
The largest cuts remove data contaminated by the 
galaxy, by the atmosphere, and by unstable offsets. Partial
descriptions of the cuts are given in Miller {\it et al.} (1999) and 
Torbet {\it et al.} (1999). In this section we describe
the cuts for the TOCO98 D-band data and the consistency checks 
as they are indicative of the process for all channels. A summary of
the cuts is given in Table~\ref{tab:cuts}. 

\subsection{Cuts to the time line}
The initial cuts are made to the raw time-ordered data in order to
excise extreme events, such as a nearby object entering the beam
or very bad weather. Data are examined 
in 6.5 second averages (24 chop averages with $f_c=3.7~$Hz). The 
internal calibration pulses are removed and a rough 
cut is made at a nominal {\it rms} level based on the long term 
observing characteristics.
This cut removes $\approx 50$\% of the data.

The next set of {\it rms} cuts is made to the synthesized data. In general, the
higher harmonics are less sensitive to atmosphere and therefore require
less severe cuts. The data are binned into fifteen minute
averages and, for each harmonic on each day, the minimum value of the
average {\it rms} is selected as the baseline. For D1, all data 
within $25\%$ of
the minimum are accepted for harmonics 8-21. For harmonics 5-7
everything within $20\%$ is accepted. For the case of D2, harmonics 8-17
are cut at $30\%$, and 5-7 are cut at $20\%$. 
Harmonics $\le 4$ are
rejected entirely as they are corrupted by atmospheric fluctuations.

We analyze 28 nights for D1 and 23 for D2. (The number is
higher for D1 due to high offsets in D2 on several
nights leading to the decision to exclude these data from the analysis.) In
order to prevent signal contamination from times
of large atmospheric fluctuations, the previous and subsequent fifteen
minute segments are eliminated from each segment cut by the above
criteria. The effect of this cut is to keep 5-10 hour blocks of continuous good
data in any day and to eliminate transitions into periods of 
poor atmospheric stability. On a typical day that passes the 
initial {\it rms} cut, an additional 40\% of the synthesized
data is removed due to atmospheric fluctuations. The net result is that
26\% of the data are kept.

As discussed above, data are placed into 768 bins in right ascension
with RA=$0\deg$ corresponding to the first bin. All data
falling between bins 288 and 555 are cut in order to eliminate
observations of the Galaxy from the CMB analysis. This is equivalent to
cutting all data with $135^{\circ}<RA<260^{\circ}$ centered at
$\delta=-60^{\circ}$.
By observing at an azimuth of $208^\circ$, the galaxy cut occurs during 
the day, when the data are not 
generally useful for CMB
observations at 144 GHz because of the atmosphere.
The Galactic cut overlaps well with the atmospheric cut. On a day 
during which there are minimal atmospheric fluctuations, we end up
cutting only $35\%$ of the data.

When the {\it rms} chopper position over a single chop deviates by more than
$0\deg.015$ from the average position, the data are cut.
This includes times when the chopper is intentionally shut
off due to testing and maintenance as well as time during high winds.
This cut eliminates $\approx 3\%$ of the data. 

The absolute timing is done through the Global Positioning System (GPS).
Data are eliminated during GPS drop-outs. There was also an error in the
first frame of each logical file which made the GPS read out incorrectly
for that frame. The first file in each set is therefore rejected
($\approx 5\%$ of the data). The files were generally started directly
following a calibration or during the day when we were at the telescope,
thus the percentage of these data which would otherwise be retained for
CMB analysis is small.

\subsection{Cuts to the binned data}

The goal of the cuts to the binned data 
is to ensure that only long periods of
uninterrupted data are included in the final analysis. 
We cut away the low density regions by removing any fundamental bin with
fewer than fifteen of the adjacent bins filled. This is done twice to
ensure that stragglers are removed. Finally, the entire file
(approximately one day) is removed if there are fewer than 100 usable
bins (approximately three hours of uninterrupted clean data). More
than 90\% of the days analyzed for CMB anisotropy 
have continuous 5-10 hour blocks of data.

\begin{table*}[t]
\caption{Cuts for TOCO98 D-band data}
\vbox{
\tabskip 1em plus 2em minus .5em
\halign to \hsize {
   #\hfil &#\hfil &#\hfil &\hfil#\hfil \cr
\noalign{\smallskip\hrule\smallskip\hrule\smallskip}
Data cuts for the season  \cr
 Total number of days analyzed\tablenotemark{a}&54~days& \cr
   Number of days used (D1) & 28~days ($52\%$)&\cr
   Number of days used (D2) & 23~days ($47\%$)&\cr
   Percentage of data kept on good day &{$53\%$} &\cr
   Net percentage kept\tablenotemark{b}& {$26\%$} &\cr
\noalign{\smallskip\hrule\smallskip}
Data cuts for a typical good day & Data Removed& \cr
Galaxy&32\% & \cr
{\it rms} (overlaps with Galaxy cut)& 24\% & \cr
Chopper& 3\% & \cr
Pointing& 1\% & \cr
GPS (overlaps with {\it rms} and Galaxy cut)& 5\% & \cr
Net cut before beam synthesis & 41\% &\cr
Net cut after beam synthesis & 45\% &\cr
Net cut to the binned data\tablenotemark{c}& 47\% & \cr
\noalign{\smallskip\hrule\vskip1pt\hrule\smallskip}
}}
\tablenotetext{a}{This is the number of consecutive days on which the
SIS system operated. After 54 days, the refrigerator cold head
malfunctioned and heated to the point where the SISs were unusable.}
\tablenotetext{b}{This is the net observing efficiency for data 
of sufficient quality for CMB analysis.}
\tablenotetext{c}{This corresponds to 53\% of data kept on a good day.}
\label{tab:cuts}
\end{table*}

\subsection{Offset Removal}

An offset is the value of a given harmonic 
when the signal on the sky is zero. 
Equation~\ref{eq:demod} gives the expression for an element of
synthesized data. Each element in
the final data file is the average value over the RA
bin,
\begin{equation}
t_n^{bin}={1\over N_{cib}}\sum_{j=1}^{N_{cib}} t_{n,j},
\label{eq:tarray}
\end{equation}
where $t_{n,j}$ is the result from each chop cycle and $N_{cib}$ is
the number of chopper cycles in a RA bin. Thus, on a given night, 
each RA bin has a single value of $t^{bin}$ for each
harmonic. These values are examined as a function of RA bin and both a
slope and a mean are removed from each harmonic on each night before the
nights are combined in order to reduce the potential effect of
variations in the offset over long timescales. The slope subtraction
has little effect on the final result; its removal is prudent though 
not essential.

Because the best data do not exactly overlap each night, a potential
bias occurs when the mean and slope are subtracted over different
regions of sky. Typical offsets
for the TOCO98 D-band data are $\approx 150\pm 75~\mu$K in absolute value
and drift over
a timescale of days. When each full night of data of one harmonic
is coadded and analyzed as a function of night throughout the campaign, 
$\chi^2/\nu$ for deviation from a flat line is between one and four. 
The high-$l$ offsets are generally more stable.
Because we keep only large sections of contiguous data
and because the offset is stable, the effect of slightly different
sky coverage per evening is negligible. In the analysis, we approximate the
slope and offset removal as removing a single mean and slope from the entire data
set as discussed below. 

The chopper-position dependent offsets  (Figure~\ref{fig:obs_offset}) 
discussed in Section~\ref{sec-offsets} are effectively filtered by 
the beam synthesis. We check this by applying the synthesis vectors to
the average chopper-synchronous signal by setting $d_i=O_i$ in
equation \ref{eq:demod}. We then examine the variation in the resulting 
synthesized offset as a function of harmonic and observing night. 
We find that the standard deviation of the resulting $O_n$ is $<20\mu$K 
for the harmonics used in the analysis. Because the offsets are
subtracted, this variation does not affect the final result.

\subsection{Ergodicity of noise}

Well defined noise properties are essential for the analysis of CMB
data. After the offset subtraction and data cuts, we grid the $t_n$ for
each harmonic with pixel number vs. night of observations. We find that 
the average error bar is independent of night and pixel number. 
For a given pixel, we also check that the variation in the mean value
from night to night is consistent with the average of the variances
separately determined each night.  

In the high signal-to-noise channels, $\delta T_l^2$ can be 
found by forming:
\begin{equation}
\delta T_l^2=(\Delta^2_{tot}-\Delta^2_{inst})/I(W)
\label{eq:tsanity}
\end{equation}
where $\Delta^2_{tot}$ is the total variance of the data for one harmonic and
$\Delta^2_{inst}$ is the average variance due to instrument noise.
For $l\leq200$, the results from this calculation are within 10\% of the 
results for the full likelihood analysis \cite{Torbet99}. At $l<150$ the
noise is potentially the most problematic because the offsets are
generally larger and the contribution from the atmosphere is larger
than for the higher $l$ data. That a simple method based on average
noise properties gives the same answer (within $1\sigma$) as the full 
likelihood analysis,
with its detailed attention to correlations and drift subtraction,
gives us confidence that the noise estimates are robust and stable.
There is nothing in the instrument or sky of which we are aware that changes 
at the temporal frequencies and spatial scales 
associated with $l>200$. Although the signal to noise is lower for large
n-pt beams, the data are
stable and insensitive to cuts. 

\section{Likelihood Analysis and Tests of Data Selection}

\begin{figure*}
\epsscale{0.75}
\plotone{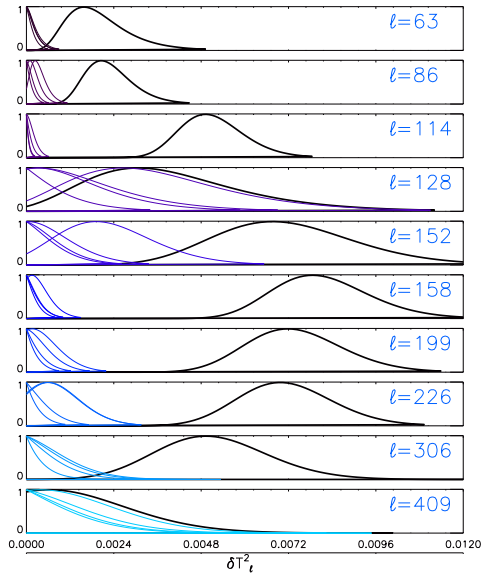}
\caption{Likelihood plots for the TOCO data. All likelihoods are
normalized to unity at the maximum. The x-axis is $\delta T_l^2$
in units of (mK)$^2$. The value on the right of each panel corresponds
to the effective $l$ for each Knox filter. The thick contours are for the data.
The thinner lines are the likelihoods for the null tests: quadrature, fast dither, 
slow dither, and first half minus second half signals. The $l=128$ D-band
channel is possibly contaminated by
the atmosphere, though the distribution of the null tests is consistent
with noise. The null test that is close to the signal is from
the slow dither and of all the null tests is
the one most different from zero. Such $\approx 1.5\sigma$ deviations
are not statistically unexpected.
This figure clearly shows the rise and fall
of a peak in the CMB angular spectrum. Furthermore, it 
shows that the null tests are consistent with zero signal. 
The D-band data are at $l=128,152,226,306,\& 409$.} 
\label{fig:likelihoods}
\end{figure*}

We find the most likely value of $\delta T^2$ as a function of
angular scale with a Bayesian likelihood analysis of 
the cleaned data. The method follows the 
analysis of the Saskatoon
data though the correlation matrices are considerably
simpler due to the observing strategy and the focal plane array.
One feature of the likelihood analysis is that
that channels and harmonics can be combined in a precise and 
unambiguous way to increase the signal-to-noise. An added benefit of the
combined bands is that correlation between combined bands is reduced over that 
for a single harmonic. The likelihoods for the HEMT data and 
SIS data are shown in Figure~\ref{fig:likelihoods}.

We test that data selection eliminates the atmospheric and
instrumental contaminants without biasing the result by
performing a set of internal consistency tests or null tests. 
Data are multiplied by several different sets of synthesis vectors that have no
sensitivity to the sky signal and the analysis is carried out in the
same way as for the data which have been multiplied by the
sky-sensitive synthesis vectors. We determine that the data have been
properly selected when the set of null tests fails to show 
a signal for individual synthesized beams. The data selection is
therefore blind to the signal on the sky,
minimizing potential bias introduced by expectations. 

The four null tests probe different time scales. The quadrature signal,
the difference in signal between the first and second half
of the chop, is sensitive to variations at 2~Hz. For the fast and slow
dither, we look at the difference between subsequent 0.5
second and 10 second averages respectively. We also examine the
difference night to night and between the first and second half of the
campaign. In the TOCO98 analysis of
individual harmonics for each channel, there was one ``failure'' 
($>2\sigma$ deviation from a null signal) out of 120 
tests, well within expectations. 
The results of some of the null tests for the full bands
are shown in  Figure~\ref{fig:likelihoods}.

\subsection{Effects of selection criteria on the data}

Following the selection of the best cuts, we relax 
the selection criteria or methods in four ways. 
For each test, we perform the full 
likelihood analysis. We emphasize that the nominal cut is based on the 
null tests, not the data. When the criteria are relaxed, the null tests
either show signs of contamination, when the likelihoods change, or are 
unaffected, when the likelihoods do not change. In the following we 
focus on the D-band data, though similar tests were performed on the
HEMT data.

Test 1: The minimum number of filled bins on a given day is
increased from 100 ($\approx 3$ hours) to 250 ($\approx 8$ hours) and
all days not common to both D1 and D2 are rejected from the
analysis. The result is a decrease in the significance of the detections
(due to permitting less data in the final analysis) but a change of less
than $1\sigma$ for the two low-$l$ points and less than $\sigma/2$ for
the three high $l$ points.

Test 2: The nominal calibration model is determined from fits
to Jupiter and the internal calibration pulses. Instead, we perform the
analysis simply using the calibration value of the closest Jupiter calibration.
The result is a change of less than $\sigma/2$, primarily due to a 
different calibration.

Test 3: We make the harmonic dependent cut to the synthesized data 
more stringent. Instead of cutting 45\% for a typical good day 
as shown in Table~\ref{tab:cuts}, 65\% is cut. The more stringent
cut is most pronounced in the lower harmonics because they are most affected by
atmospheric contamination. The cut results in upper limits
for the $l=128$ and $l=158$ points, consistent with the
decreased amount of data. Above the 7-pt, the amount of data cut and the
resulting likelihoods change very little. We also perform this test
by eliminating just 50\%  (as opposed to 45\%) with similar results.

Test 4: We apply only the initial cut to the raw data and so
on a typical good day, 41\% of the data are kept (before the Galaxy cut).  
This has the effect of introducing some
atmospheric contamination. The result is that the  $l=128$ and $l=158$
points show an increase in signal of $0.8\sigma$ and $1.3\sigma$ respectively
while the data points corresponding to the highest three groups of harmonics
show no significant change. 

We have emphasized the TOCO98 D-band data because the points
at $l=128$ and $l=152$ are more sensitive to the cuts than all the 
other data. This is due to the large atmospheric opacity at 144 GHz
and large angular scales. The nominal cut corresponds to the best null tests
and corresponds roughly to the sections of raw data one would
select by eye. The D-band data for $l>200$ are insensitive to the 
cuts as are all of the HEMT data.

\subsection{Correlations}

The off-diagonal terms in the noise correlation matrix, $C_N$, are small.
The atmospheric correlation coefficients between channels (Q and
Ka band) for TOCO97 and between D1 and D2 for TOCO98 are of order
0.05. We examine the autocorrelation function of the data for individual
harmonics and find negligible correlations between RA bins (due to
atmospheric fluctuations). The largest off-diagonal terms in {\bf M}
are $<0.4$ and are in $C_T$. As part of our checks, we ensure that
the likelihood is stable to changes of $\approx~30~$\% the values of 
the off-diagonal terms. 

\section{Tests for Systematic Effects}
\label{sec-systematics}

There are potential sources of systematic error that would not 
be revealed by the tests described above. These sources of error,
along with their maximum contribution to the data, are summarized 
in Table~\ref{tab:systematics} and described below.
Again, we focus on the D-band data. We note that no effect 
significantly affects the final results and in particular no effect
can produce the decrease in power that is observed between $l\approx
200$ and $l\approx 400$.

The systematic effects involve either beam smearing
(Sections~\ref{sec-chop_jitter} - ~\ref{sec-beam_wrong}) or
another mechanism (Sections~\ref{sec-mis_cal_amp} -
~\ref{sec-phase_offset}). 
For those effects related to beam
smearing, the uncertainties are found through computing
$I(W)$ for the smeared beam.
The fractional error for an individual harmonic is given by
\begin{equation}
{\Delta(\delta T_l)\over \delta T_l}={1\over\delta T_l}{\partial(\delta
T_l)\over\partial\sqrt{I(W)}}={\delta \sqrt{I(W)}\over\sqrt{I(W)}}.
\label{eq:error_prop}
\end{equation}
The fractional error
in groups of harmonics is smaller because the errors are computed 
for the highest harmonic in a band group,
the one with the fewest number of physical beams between nulls in the
synthesized beam and therefore the
one most affected by beam smearing. 

Errors are computed for D1, unless otherwise noted,
because it has a smaller beam than D2 and is consequently
more affected by smearing. Nominal values of $\sqrt{I(W)}$ are the following:
$\sqrt{I(W)}=0.377$ for D1 and $\sqrt{I(W)}=0.350$ for D2 for the case
of the 16-pt and 17-pt beams respectively.

\begin{table*}
\vbox{
\tabskip 1em plus 2em minus .5em
\halign to \hsize {
   #\hfil &\hfil#\hfil \cr
\noalign{\smallskip\hrule\smallskip\hrule\smallskip}
   Systematic Effect& Maximum Resultant Change in $\delta T/T$ \cr
\noalign{\smallskip\hrule\smallskip}
Chopper jitter & \hfill$< 0.1$\% \cr
Chopper zero offset drift & \hfill$< 0.2$\% \cr
Miscalibration of chopper amplitude & \hfill$< 3$\% \cr
Mis-calibration of azimuth & \hfill$< 0.1$\% \cr
Mis-calibration of elevation & \hfill$< 1$\% \cr
Incorrect determination of beam size & \hfill$< 1$\% \cr
Electronic roll-off (between data points)& \hfill$<0.2$\% \cr
Gain variation in offsets & \hfill$<3$\% \cr
Chopper phase offset & \hfill$<0.5$\% \cr
\noalign{\smallskip\hrule\vskip1pt\hrule\smallskip}
}}
\caption{Systematic Effects}
\label{tab:systematics}
\end{table*}

\subsection{Chopper Jitter}
\label{sec-chop_jitter}
Jitter in the chopper position, due to wind or
electronic malfunction, smears the beam.
All data for which the {\it rms} of the chopper
position deviates more than
$0.015\deg$ from the nominal position are rejected from the
analysis. The maximum resulting error in either D1 or D2 in any harmonic
(resulting from an increase in beam size in the direction parallel to
the chop of $0.015\deg$)
is $\delta\sqrt{I(W)}=-3.0\times 10^{-4}$, or $\delta T_l/T_l\approx 0.1\%$. 

\subsection{Chopper Zero Offset Stability}
\label{sec-chop_zero}
Beam smearing can be caused by a drift in the electronic zero point
of the chopping mirror. The zero point is monitored
throughout the campaign. We find that from the beginning to the end of
the analyzed data the zero drifts by $\approx 0\deg.03$. We can place an
upper limit on the extent to which this can affect our result by
considering a smearing of the beam in the direction parallel to the chop
of $0\deg.03$. This results in $\delta\sqrt{I(W)}=-7\times 10^{-4}$, or $\delta
T_l/T_l\approx 0.2\%$. 

\subsection{Mis-calibration of Azimuth}
\label{sec-mis_cal_az}
If the azimuth changed over a period of time,
the true beam could be uncertain by an amount equivalent to this
change. The relative error in azimuth is $<0\deg.01$ which could manifest
itself as an increase in beam size in the direction parallel to the chop
by a maximum of $0\deg.01$. The resultant errors are
$\delta\sqrt{I(W)}=-2.0\times 10^{-4}$, or $\delta T_l/T_l< 0.1\%$.

\subsection{Mis-calibration of Elevation}
\label{sec-mis_cal_el}
Similarly, if the elevation changed over time,
the true beam could be uncertain by an amount equivalent to this
error. The relative error in elevation is $<0\deg.01$ which could manifest
itself as an increase in beam size in the direction perpendicular to the chop
by a maximum of $0\deg.01$. The resultant errors are
$\delta\sqrt{I(W)}=-3\times 10^{-3}$, or $\delta T_l/T_l\approx 0.8\%$.

\subsection{Incorrect Determination of the Beam}
\label{sec-beam_wrong}
If the beam was incorrectly measured, there would be an
effective beam smearing in either the direction parallel or
perpendicular to the chop. Errors in the $\theta_{FWHM}$ of the 
beam are $\approx 0\deg.005$ and
$\approx 0\deg.008$ for D1 and D2 respectively. The
maximum resulting
error in either channel in any harmonic is $\delta\sqrt{I(W)}=2.5\times
10^{-3}$, or $\delta T_l/T_l\approx 0.7\%$. This is an upper
limit because errors in the determination of the beam affect the
calibration as well and are therefore partially compensated.

\subsection{Mis-calibration of Chopper Amplitude}
\label{sec-mis_cal_amp}
If the chopper amplitude were mis-calibrated, the assumed number
of physical beams fitting into a given synthesized beam could be
wrong. We know the amplitude of the chopper to $<2\%$ uncertainty
($2\sigma$). If the
value used in the analysis were wrong by this amount, we find the
resulting change,  $\delta T_l/T_l<3\%$ for the worst case
harmonic. Most  harmonics show a change of $\delta T_l/T_l<1\%$. The
entire spectrum also shifts a small amount in $l$. For a $2\%$ increase
in the chopper amplitude, we find $\approx 2\%$ shift in each point to
lower $l$ values. The sign of this effect is reversed if the chopper
amplitude is smaller than the assumed value.  

\subsection{Electronic Roll-off}
\label{sec-electronic_rolloff}

The likelihood contours shown in Figure~\ref{fig:likelihoods}
correspond to different post-detection frequencies as well as different
$l$.  The point at $l=226$ contains data from the 6-pt to 7-pt beams in D1 
and the 6-pt to 9-pt beams in D2 corresponding to 
$f\approx 25$~Hz. The point at $l=409$
contains the 12-16-pt beams in D1 and 12-pt to 17-pt beams in D2, 
corresponding to $f\approx 50$ Hz. By calculation and measurement
we show the electronic transfer function changes by
$< 0.2\%$ over this range and thus does not affect
the results.

\subsection{Gain Variation in the Offsets}
\label{sec-gain_var_offset}

Variations in the cryogenic temperatures and
the temperature of the warm electronics box lead to variations
in the gain. If the effects of the decrease in gain
of the warm amplifiers, the decrease in sensitivity of the SIS mixers,
and the increase in output of the calibration source with increasing
temperature are combined during the period of CMB
observations, there could be an undetected 
gain drift of $\approx 1\%$. In addition, there could be other gain 
effects of order 3\% that would escape detection because the variation
in atmospheric opacity would mask their signature in the cal pulse.
Because of the large heat capacity, such variations would occur on the 
time scale of hours.

Changes in gain affect the offset. Since the offsets are different for
different synthesized beams, it is possible to affect
the spectrum through a gain change.
When the average offsets for the combined harmonics are
multiplied by a gain variation of $\approx 3\%$, a signal with
amplitude $\approx 3\mu$K at low harmonics and $< 2\mu$K at higher
harmonics results. Not only is this signal small, but it is largely
accounted for through the offset and slope subtraction.

\subsection{Chopper Phase Offset}
\label{sec-phase_offset}

If the phase of the chopper were to drift, the sky signal would be
smeared because the beam would not be positioned according to the 
nominal chopper template.
A cut is made that eliminates from the analysis all data for which the
chopper amplitude differs from the {\it rms} average value by more than
$0.015\deg$. This corresponds to a phase shift of 0.5 samples.
A shift of this size contributes $<0.5~\mu~$K in $\delta T_l$.

\section{Foreground emission and Results}
\label{sec-foregrounds}

\begin{table*}[ht]
\caption{Summary of Foreground Contributions\tablenotemark{a}}
\small{
\vbox{
\tabskip 1em plus 2em minus .5em
\halign to \hsize {
     #\hfil &\hfil#\hfil &\hfil#\hfil &\hfil#\hfil &\hfil#\hfil &\hfil#\hfil \cr
\noalign{\smallskip\hrule\smallskip\hrule\smallskip}
Foreground & $l=60$\tablenotemark{b}  & $l=85$ &
$l=115$ & $l= 150$ & $l=200$\cr
\noalign{\smallskip\hrule\smallskip}
IRAS sources at 30 GHz ($\mu$K) & $\cdots$ & 19 &$\cdots$ & $\cdots$ &$\cdots$ \cr
Radio sources at 30 GHz ($\mu$K) & $\cdots$ & 16 & 23 & 16 & 13 \cr
Radio sources at 40 GHz ($\mu$K) & $\cdots$ & 11 & 10 & $\cdots$ \cr
SFD 30 GHz ($\mu$K)  & 18 & $\cdots$ &$\cdots$ & $\cdots$ &$\cdots$ \cr
SFD 40 GHz ($\mu$K) & $\cdots$ & 9 &$\cdots$ & $\cdots$ &$\cdots$ \cr
\noalign{\smallskip\hrule\vskip1pt\hrule\smallskip}
}}}
\tablenotetext{a}{We do not give any value if the fitted foreground
contribution is $<7~\mu$K. At $l=60$ this corresponds to 
a 1\% correction and at $l=200$ a 0.5\% correction. }
\tablenotetext{b}{The foreground contribution for a typical
value of $l$.}
\label{tab:foregrounds}
\end{table*} 

\begin{figure*}
\epsscale{0.8}
\plottwo{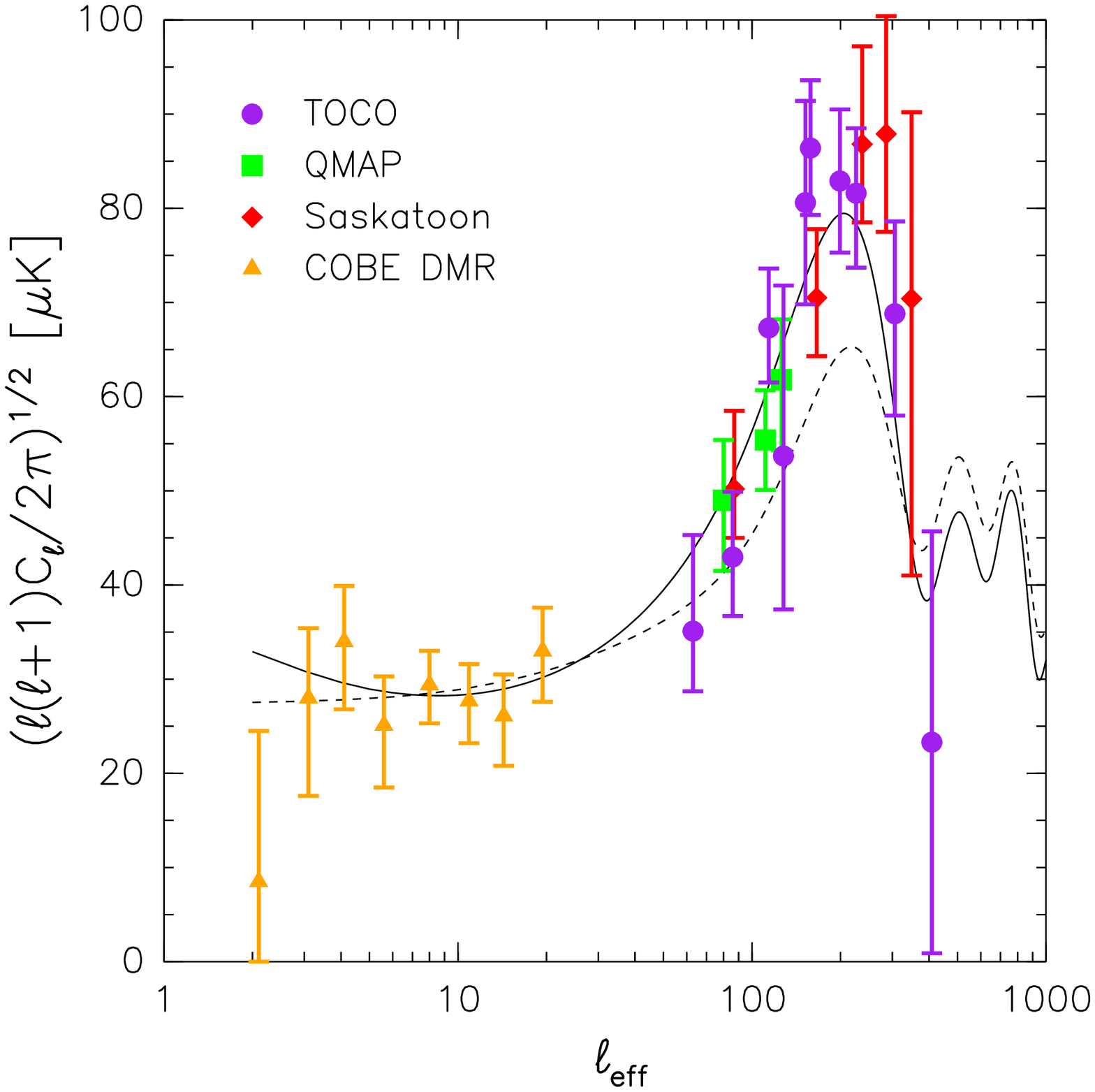}{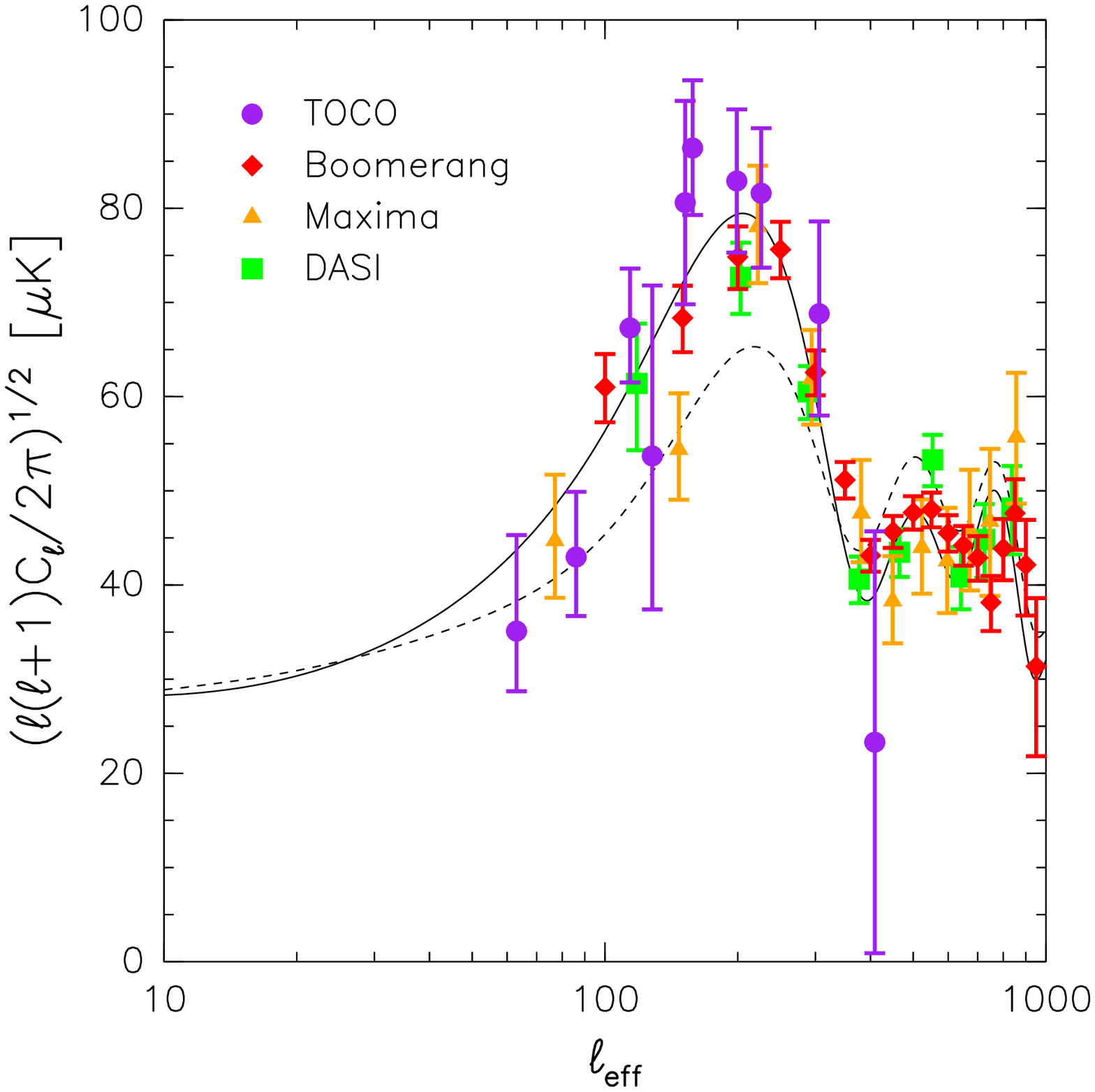}
\caption{The combination of the SK, QMAP, and TOCO data with all
corrections. Following Netterfield {\it et al.} (1997), we plot the
position of the maximum of the likelihood with error bars that 
encompass 68.3\% of the likelihood for all points where
there is a significant detection. For the last TOCO point we plot
a representation of $1\sigma$ based on Bond {\it et al.} (2000).
The plot is made with $\delta T$ as opposed to  $\delta T^2$
because $\delta T$ directly shows the signal-to-noise. 
For example, $\delta T=5\pm 1~\mu$K has a ${\rm S/N=5}$ whereas
the equivalent  $\delta T = 25\pm 10~\mu$K has 
an apparent ${\rm S/N\approx 3}$. The models are ``standard CDM''
(lower) and the best fit from Jaffe {\it et al.} (2000) (upper).
Calibration error is not included.}
\label{fig:results}
\end{figure*}

After the raw data have been reduced and binned on the sky, we 
determine the foreground emission contribution from our galaxy
and from other galaxies.  The foreground contribution at these frequencies, 
galactic coordinates, and angular scales is small 
\cite{teg00,Coble,Masi01}.
This is clear from Figure~\ref{fig:results}. The amplitude of the first
peak is measured to be roughly the same from 30 to 144 GHz when the flux is 
expressed as changes in a 2.725~K thermal emitter. The spectral index of
the fluctuations is $\beta_{CMB} = {\rm ln}(\delta T_{144}/\delta
T_{36.5})/{\rm ln}(144/36.5) = -0.04\pm 0.25$, where $\delta T_{144}$ 
is the weighted mean of the two highest points for the D-band data
and $\delta T_{36.5}$ is a similar quantity for the HEMT data 
(36.5 GHz is the average HEMT frequency).
For the CMB, $\beta_{CMB} = 0$. For dust,
$\beta_{RJ} = 1.7$ corresponds to $\beta_{CMB} = 2.05$; for free-free
emission $\beta_{RJ} = -2.1$ corresponds to $\beta_{CMB} = -1.75$.

Fits to foreground templates were not done for Torbet {\it et al.}
(1999) and Miller {\it et al.} (1999) and there was the possibility that 
the amplitude of the peak had a contribution from foreground 
emission \cite{Page00,kp00}. However, the mean frequency spectrum of the peak
clearly singles out the CMB as the dominant source of the fluctuations.
In addition, the angular spectra of the foreground
emission is much different than that of measured angular spectrum of the
CMB \cite{Miller99}.

To quantify the foreground emission, we fit to four templates:
the SFD dust map \cite{SFD} ($T_{SFD}$), the Haslam synchrotron map
\cite{Haslam82} ($T_{H}$), and the radio ($T_{r-pt}$) and IRAS ($T_{ir-pt}$)  
source compilations from the WOMBAT compilation (WOMBAT 2001). 
We have not yet fit to 
the H$_\alpha$ maps that trace microwave free-free emission. However, we note
fits to SK \cite{Gaustad96,Simonetti96} found that the
free-free contribution was negligible. In addition, cross-correlations 
between the
WHAM H$_{\alpha}$ maps \cite{Haffner99} and other CMB maps, including
QMAP, do not show a significant contribution \cite{dOC01}. 

There is a correlation between CMB maps at frequencies $<90~$GHz and 
dust maps \cite{Kogut96,dOC97,Leitch97}.  The mechanism responsible for the 
correlation is not yet certain \cite{dOC01} though 
spinning dust grains is a strong candidate \cite{Jones97,DL98}.
The dust-correlated foreground component is larger than the 
free-free component traced by H$_{\alpha}$ between 20 and 40 GHz
and is not correlated to the Haslam map \cite{dOC98b}. Thus, the 
SFD map is a good tracer of foreground emission for the HEMT data.
%
%
The galactic latitude of the CMB scan covers between $b=-8^{\circ}$,
$l=280^{\circ}$ and  $b=-34^{\circ}$,
$l=335^{\circ}$ as shown in Figure~\ref{fig:skyscan}.
Our sky scan passes near the Large Magellanic Cloud (LMC,
$b=-32\fdg9$, $l=280\fdg5$) and we remove data near it.
During the day,
we observe known sources in the galactic plane \cite{Puchalla01}.

The template fitting procedure is
restricted to the section of sky analyzed for CMB anisotropy.  
The goal is to assess the contribution to the CMB results rather
than to characterize foreground emission.
First, the template map is multiplied by the synthesized beam in a
manner such that the full CMB observing strategy is reproduced.
Except in the case of the Haslam map for the  $l>200$ D-band data
(where no correlation is expected), 
the size of the synthesized beam lobes is 
larger than the resolution of the template map. Next, for each harmonic,
we fit to the template with

\begin{equation} 
\chi^2=\sum_{k=1}^{N_{n,bins}}\bigl(t_{k}-(a_0 T_{r-pt} +\nonumber\\
a_1 T_{ir-pt} + a_2 T_{H} + a_3 T_{SFD})\bigr)^2w_{k}~~~~~~~~
\label{eq:fg}
\end{equation}
where $N_{n,bins}$ is the number of filled bins around the sky for the $n-pt$ 
function ($<N_{\rm RA}$, $t_{k}$ is the CMB temperature 
(equation~\ref{eq:tarray}), 
and $w_{k}$ is the statistical weight of each measurement. 
The fits are performed as a function of angular scale 
and frequency. Where the fit result is significant, as shown in 
Table~\ref{tab:foregrounds}, we subtract 
the foreground component in quadrature from $\delta T_l$ as 
determined from the likelihood analysis\footnote{This simple treatment,
where we ignore chance alignments of the CMB with the foregrounds
[e.g., \cite{dOC98}] and subtract the foreground in quadrature,
is sufficient because the foreground contribution to the CMB $<4$\% in
all cases. Future work will address the full treatment and the 
galactic plane.}.

The fit results may be summarized as follows: 1) The dominant source of 
contamination
at 30 and 40 GHz is from the radio point sources as traced by the 
4.85 GHz PMN catalog \cite{Griffith93,Condon93,Wright94} 
on which the WOMBAT maps are based. The extrapolation to 30
GHz is known to be problematic because the source spectra vary from
source to source \cite{Puchalla01}. We
cannot rule out a contribution from sources that are not traced
by the extrapolated PMN survey but, because of the measured spectrum of
the peak, such a contribution cannot be large.
Over our sky coverage $\approx600~$deg$^2$, we account for 
$\approx100$ sources. Most of the 
contribution comes from the largest 10\% and is spread throughout the 
observing region.
2) Contributions from synchrotron emission are negligible unless the 
synchrotron spectral 
index varies so much that the Haslam map is not a good template.
The largest {\it rms} of the Haslam map, after applying the beam synthesis, 
is 0.6~K. When extrapolated to 30 GHz with a spectral index of 
$\beta_{RJ}=-2.7$, one gets $5.5~\mu$K, consistent with the fitted values.
The fitted dust contribution
at the 30 GHz $l=60$ and 40 GHz $l=80$ points, is small but consistent
with the picture of 30 GHz dust-correlated emission. 
3) There is no significant
contamination at 144 GHz from either dust or point sources. 

Table~\ref{tab:allresults} gives the results from TOCO along with the revised
results from QMAP and SK\footnote{In addition to
some authors on this paper, SK was analyzed by Ed Wollack and Norm
Jarosik.} based on foreground analyses and new information on the calibration.
The final results for all three experiments are shown in
Figure~\ref{fig:results} along with a comparison to recent experiments.

\begin{table*}[t]
\caption{\small Summary of results from SK, QMAP, and TOCO.}
\small{
\vbox{
\tabskip 1em plus 2em minus .5em
\halign to \hsize {
     #\hfil &#\hfil &#\hfil &#\hfil &#\hfil &#\hfil &#\hfil &#\hfil \cr
\noalign{\smallskip\hrule\smallskip\hrule\smallskip}
Experiment & 
$l_e$\tablenotemark{a} &
$N_{\rm n-pt}$\tablenotemark{b} & 
${\rm N_{bins}}$\tablenotemark{c} & 
$\delta T_l^{orig~~}$ \tablenotemark{d} & 
$\delta T_l^{fc~~}$\tablenotemark{e} &  
$\delta T_l$\tablenotemark{f} &   
$\delta T_l^2$ \cr
\omit & & & & ($\mu$K) & ($\mu$K) & ($\mu$K) & ($\mu$K)$^2$ \cr 
\noalign{\smallskip\hrule\smallskip}
TOCO  & $63^{+18}_{-18}$  & 2       & 16,20   & $39.7^{+10.3}_{-6.5}$&
      35.4 & $35.1^{+10.2}_{-6.4}$& $1232^{+820}_{-408}$  \cr
QMAP  & $80^{+41}_{-41}$  &$\cdots$ &$\cdots$ & $47^{+6}_{-7}$ & $46.0$
      &   $48.3^{+6.4}_{-7.5}$ & $2401^{+668}_{-679}$\cr
TOCO  & $86^{+16}_{-22}$  & 3       & 17-28   & $45.3^{+7.0}_{-6.4}$ &
      43.4& $43.0^{+6.9}_{-6.3}$ & $1846^{+644}_{-499}$ \cr
SK    & $87^{+39}_{-27}$  & 15      & 24,48   &  $49^{+8}_{-5}$ & 48.0
      & $50.2^{+8.3}_{-5.2}$ & $2520^{+902}_{-495}$  \cr
QMAP  & $111^{+64}_{-64}$ &$\cdots$ &$\cdots$ & $52^{+5}_{-5}$ &
      $52.0$ & $54.6^{+5.3}_{-5.3}$ & $3069^{+615}_{-559}$   \cr
TOCO  & $114^{+20}_{-24}$ & 6       & 22-42   & $70.1^{+6.3}_{-5.8}$ &
      68.0 & $67.3^{+6.3}_{-5.8}$ & $4529^{+888}_{-747}$ \cr
QMAP  & $126^{+54}_{-54}$ &$\cdots$ &$\cdots$ &  $59^{+6}_{-7}$ &
      $58.0$ & $60.9^{+6.4}_{-7.5}$ & $3819^{+832}_{-871}$  \cr
TOCO  & $128^{+26}_{-33}$ & 1       & 84      & $54.6^{+18.4}_{-16.6}$ &
      54.6 &  $53.7^{+18.1}_{-16.3}$  & $2884^{+2272}_{-1485}$ \cr
TOCO  & $152^{+28}_{-38}$ & 3       & 84      & $82.0^{+11.0}_{-11.0}$ &
      82.0 &  $80.6^{+10.8}_{-10.8}$  &   $6497^{+1858}_{-1625}$      \cr
TOCO  & $158^{+22}_{-23}$ & 6       &29-55         & $88.7^{+7.3}_{-7.2}$ &
      87.2 &  $86.4^{+7.2}_{-7.1}$  &   $7465^{+1296}_{-1177}$  \cr
SK    & $166^{+30}_{-43}$ & 10      & 48,96   & $69^{+7}_{-6}$ & 67.6
      & $70.5^{+7.3}_{-6.2}$ & $4970^{+1080}_{-836}$  \cr
TOCO  & $199^{+38}_{-29}$ & 11      & 54-82     & $84.7^{+7.7}_{-7.6}$ &
      83.7 & $82.9^{+7.6}_{-7.6}$  &  $6872^{+1318}_{-1202}$ \cr
TOCO  & $226^{+37}_{-56}$ & 6       & 125     & $83.0^{+7.0}_{-8.0}$ &
      83.0 &  $81.6^{+6.9}_{-7.9}$  &  $6659^{+1174}_{-1094}$    \cr
SK    & $237^{+29}_{-41}$ & 4       & 48,96   & $85^{+10}_{-8}$ & 83.3
      & $86.8^{+10.4}_{-8.3}$ & $7535^{+1914}_{-1372}$  \cr
SK    & $286^{+24}_{-36}$ & 4       & 48,96   & $86^{+12}_{-10}$ & 84.3
      & $87.9^{+12.5}_{-10.4}$ & $7726^{+2354}_{-1720}$  \cr
TOCO  & $306^{+44}_{-59}$ & 6       & 165     & $70.0^{+10.0}_{-11.0}$ &
      70.0 &   $68.8^{+9.8}_{-10.8}$   &  $4733^{+1445}_{-1369}$  \cr
SK    & $349^{+44}_{-41}$ & 5       &48,96    & $69^{+19}_{-28}$ & 67.6
      & $70.4^{+19.8}_{-29.1}$ & $4956^{+3180}_{-3275}$ \cr
TOCO\tablenotemark{g}  
      & $409^{+42}_{-65}$ & 9       & 250     & $<67~(95\% conf)$ & 
      \dots & $23.3^{+22.4}_{-22.4}$ & $545\pm2043$ \cr
\noalign{\smallskip\hrule}
}}}
\tablenotetext{a}{$l_e$ is computed from the window function or the Knox
filter following Bond (1994).  In practice, we find little difference
between the combined weighted windows \cite{net97} and 
Knox filters \cite{Knox99}.}
\tablenotetext{b}{The total number of individual n-pt functions
combined in the covariance matrix {\bf M}. We emphasize that
all known correlations are accounted for.
For QMAP, n-pt functions correspond to different eigenmodes for the map.}
\tablenotetext{c}{The number of data points or RA bins. When two
numbers or a range are given, not all n-pt beams have the same RA bins.}
\tablenotetext{d}{The originally published values following the 
convention in Netterfield {\it et al.} (1997). Calibration error
is not included.}
\tablenotetext{e}{The original value corrected for foreground
emission ($fc$) . For SK, \cite{dOC97} found a $\approx 2$\% contamination of the 
data due to a foreground component correlated with dust emission.
For QMAP a similar correction was found \cite{dOC99}. The entries
show the spectrum after the corrections. The corrections are done
separately for Ka and Q bands before they are combined.}
\tablenotetext{f}{The foreground and calibration corrected values. The 
recalibration is based on new information about the 
calibration sources or, in the case of TOCO, on a 1-2\% correction
for the electronic bandpass. After SK data were published, \cite{Mason00} reported
a more accurate calibration of Cas-A, which we used for QMAP. This led
to an increase of 4\% for the SK data and a reduction in the calibration
error from 14\% to 10\%. There is also a small correction to put the
results in the $l(l+1)/2\pi$ format as opposed to the 
$l(2l+1)/4\pi$. Calibration error is not included.}
\tablenotetext{g}{Originally, the 95\% upper limit and likelihood
curves were given for this bin whereas $1\sigma$ error bars were plotted
for the other points. We here adopt the 
convention in Mauskopf {\it et al.} (2000), and give all results 
with  $1\sigma$ error bars. The data are the same as before \cite{Miller99} 
except for the 1.7\% calibration correction.
The likelihood distribution is not Gaussian. The value of $23\pm22~\mu$K
matches the distribution in the sense that the likelihood peak is at
$23~\mu$K and $\approx95$\% 
of the probability is $<67=23+2\sigma~\mu$K.
The value of $545\pm2043~(\mu {\rm K})^2$ comes from fits to the 
likelihood following Bond {\it et al.} 2000 and is often used to
represent the likelihood. Note that $\sqrt{545+2\sigma}=68~\mu$K.
We set the error bar on $\delta T_l$ by forming
$(\sqrt{545+2\sigma}-\sqrt{545})/2.~\mu$K.
For detailed analyses, the full likelihood as shown in Figure 9 should 
be used.}
\label{tab:allresults}
\end{table*}
%
%

\section{Discussion}

Since the discovery of the anisotropy \cite{smoot92}, there have been
many CMB anisotropy experiments at $l<1000$ in addition to the ones noted
thus far [ARGO \cite{dB94,Ratra99a}, MSAM \cite{Wilson00}, 
UCSB/SP \cite{gunder94},
White Dish \cite{Ratra98}, Python \cite{Ruhl95}, BAM \cite{Tucker96}, 
IAB \cite{pc93}, Tenerife \cite{Hancock94},
MAX \cite{Lim96}]. The SK experiment was the first to show the rise of the 
CMB angular spectrum in the region between $l=50$ and $l=250$
independent of any other experiment. 
At the time, the amplitude of the peak was considered high 
though consistent with the subsequently
popular $\Lambda$-CDM models. Until TOCO, the SK spectrum was 
in mild conflict with other experiments.

Though a straightforward read of the data prior to TOCO
suggested a peak at $l\approx 200$ [e.g., SK and OVRO \cite{Leitch98} or
SK and CAT \cite{bak99}] there were lingering questions
of cross calibration and point source subtraction. In fact,
if one did not include the SK data, the 
combined analysis of UCSB/SP, ARGO, MAX, White Dish, 
and SuZIE favored an open universe \cite{Ratra99b}.
Dodelson and Knox (2000) showed that TOCO alone singled out a
flat universe as the best model. Others \cite{bahcall99,bjk00}, including
Dodelson and Knox, showed that the combination of all the data singled
out a flat universe.

QMAP was designed to measure the anisotropy by direct mapping.
Degree-resolution high signal-to-noise maps were made 
that could be compared to each other and, because they covered the 
same region of sky, confirmed the $l<150$ Saskatoon results \cite{Xu00}.
The strategy of direct total power mapping was later employed
by the BOOMERanG and MAXIMA experiments, which ushered in 
very high signal-to-noise sub-degree resolution mapping. 
 
The TOCO experiment showed both 
the rise and fall of the first peak and showed its spectrum was
that of the CMB with a single instrument
(Figure~\ref{fig:results}). These results have
since been amply verified by BOOMERanG, MAXIMA, and DASI.
The rise to the maximum has high signal-to-noise; 
the fall for $l>200$ is also clearly evident
though worth quantifying. Miller {\it et al.} (1999) reported 
that the 95\%
upper limit of the last point in D-band, $\delta T_{409}$, was 
just below the $2\sigma$ lower bound of the point of the peak, 
$\delta T_{226}$. The penultimate point
was not included in the assessment.
Here we perform a more complete analysis.
We focus on just the last three D-band points, $\delta T_{226}$, 
$\delta T_{306}$, $\delta T_{409}$,  because the calibration
uncertainty is common to all and drops out of the analysis.
There are two ways a peak could be detected:
$\delta T_{226}> \delta T_{409}$, and $\delta T_{306}> \delta T_{409}$
but with $\delta T_{409}\le \delta T_{226}$.
The net probability that a peak has been detected is given by
the sum of the probabilities of these two possibilities.

%
%
%
%
%
%
%

We Monte Carlo the likelihood distributions to 
determine the above probabilities and to
investigate the effects of correlations.
We find $P(\delta T_{226}> \delta T_{409}) = 0.99614$ 
and  $P(\delta T_{306}> \delta T_{409}|\delta T_{409}\le \delta T_{226}) =
0.00369$. Thus, the net probability that a peak has been detected is
0.99983, or loosely speaking a $>3.7\sigma$ detection. The 
correlations among these three points are positive and small,
of order 0.01. When the correlations are accounted for,
the net effect is to increase the significance of a detection
of a peak. We emphasize that the detection is model independent,
calibration independent, and conservative in the sense that
if there is slight contamination by point-sources
or there is some undetected source of
correlation, the probability of the detection {\it increases.} 

There are now many analyses that extract cosmological parameters
assuming that the models from CMBFAST \cite{selzal} 
describe Nature, [e.g., \cite{bartlett98,line99,tegzal,Jaffe01}]. Here 
we compute $\delta T_{peak}$ and $l_{peak}$ directly from the data.
This allows parameter estimation with the minimal amount of model 
dependence. From the TOCO and SK data, the average amplitude 
of the peak between $l=160$ and $l=240$ is 
$\delta T_{peak}=80.9\pm3.4\pm5.1~\mu$K where
the first error is statistical and second error is for calibration 
(the $1~\mu$K shift from \cite{Miller99} is mostly due to the 
new calibration). Following  
Knox and Page (2000), we find $l_{peak}=216\pm14$
in a relatively model independent parametrization
\footnote{The peak value from this method, $\delta T_{peak}=86\pm8~\mu$K,
is sensitive to the type of fit whereas $l_{peak}$ is relatively
fit-independent. We prefer the straightforward average as it is computed
directly from the data.}.

The SK, QMAP, and TOCO experiments used a variety of techniques,
separate data reduction and analysis pipelines, and two different calibrators.
These different experiments, rich with consistency checks,
trace out a peak.
We have described the systematic checks, focusing on the TOCO data,
and have not found any instrumental effect or data reduction artifact
that could mimic
or produce the signal we see. It is possible that extragalactic 
sources with spectra different from the ones we assume 
could alter our results but the effect would be small and
accounting for it would tend to enhance the downturn for $l>220$. 
In conclusion, these experiments,
in particular the TOCO experiment, have measured the rise, amplitude, 
position, and fall of the first peak in the angular spectrum of the CMB.

\begin{figure*}
\epsscale{0.7}
\plotone{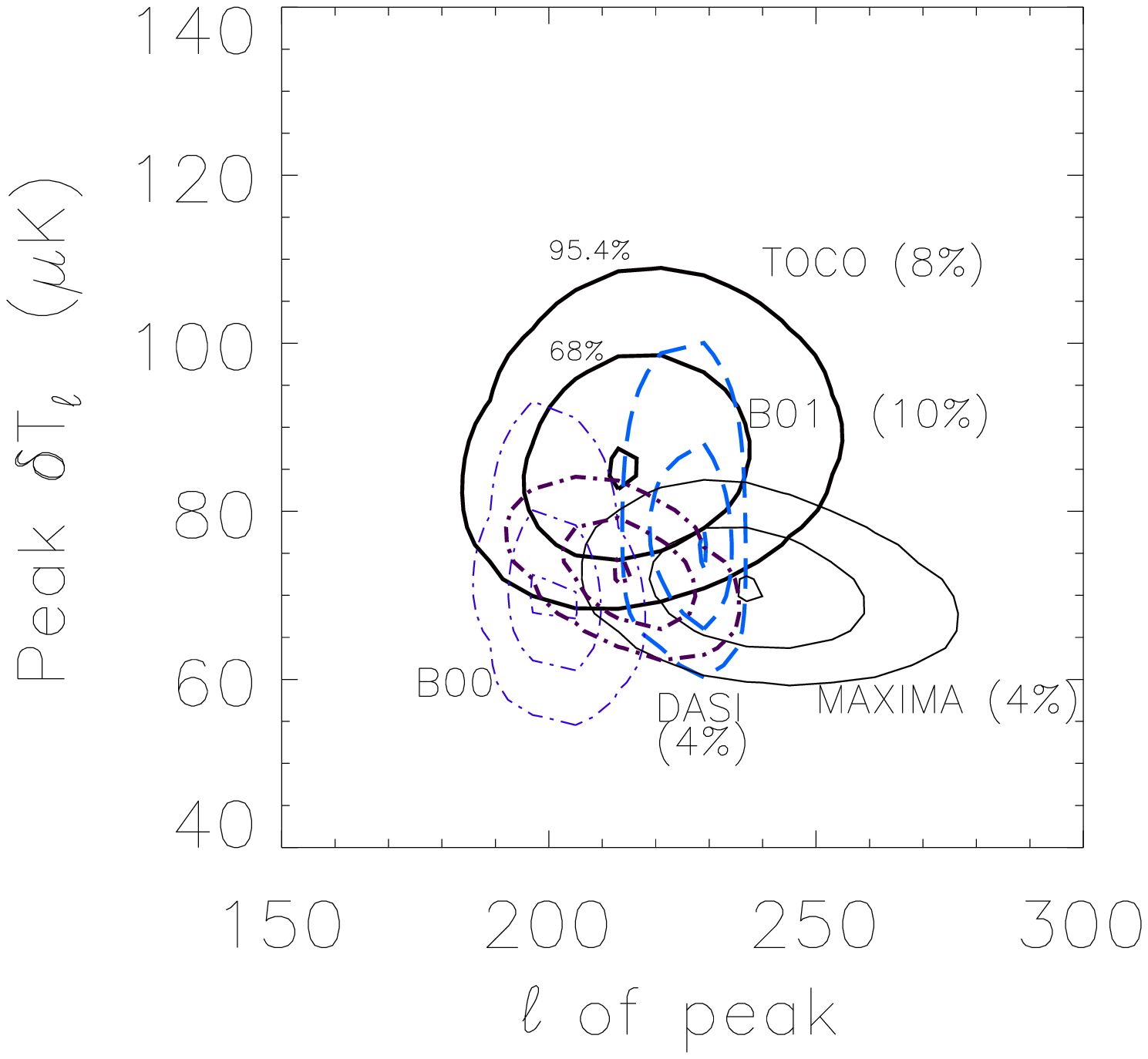}
\caption{We show the position and amplitude of the first peak 
following the Gaussian Temperature method \cite{kp00}. For TOCO  
$l_{peak}=216\pm14$ and $\delta T_l=86\pm8~\mu$K. These
values are slightly different than those reported before
($l_{peak}=212$ and $\delta T_l=88~\mu$K) because the of 
the calibration and foreground corrections.
The preliminary BOOMERanG
\cite{dB00} data give $l_{peak}=201$ and $\delta T_l=70~\mu$K; 
the reanalyzed and expanded data set \cite{net01} gives $l_{peak}=226$
and $\delta T_l=77~\mu$K. (We omit the North American Flight.) 
The BOOMERanG beam error is not accounted for; it will
tend to broaden the distribution in $l$. MAXIMA \cite{Lee01} yields 
$l_{peak}=238$ and $\delta T_l=71~\mu$K and DASI \cite{Halverson01} 
gives $l_{peak}=213$ and $\delta T_l=74~\mu$K. These values are close ($<1\sigma$)
to the values found in de Bernardis {\it et al.} (2001) using different 
methods. Calibration
error has been taken into account though the correlations between
bands have not. When treated consistently, TOCO, B01, MAXIMA, and
DASI pick out values for the peak position and amplitude that
are within $2\sigma$ of each other.}
\label{fig:peak_results}
\end{figure*}

\section{Acknowledgements}

The QMAP and TOCO experiments took place over six years and 
involved many colleagues. 
We gratefully acknowledge conversations with and help from 
Chuck Bennett, Joe Fowler, Lloyd Knox, Steve Meyer, Steve Myers,  
Bharat Ratra, David Spergel, Suzanne Staggs, and 
Dave Wilkinson. The TOCO and QMAP experiments
made ample use of the insights and previous efforts of Norm Jarosik and 
Ed Wollack. As just one example, Norm's electronics have worked flawlessly 
for six years.
Max Tegmark and Angelica de Oliveira-Costa led the science analysis
of the QMAP data and have had a large influence on the work
presented here. The Princeton Machine Shop time and again
came up with rapid and creative solutions to our mechanical problems.
Operating in Chile would have been considerably more difficult
without the kind and frequent help of Angel Ot\'arola.
The Cerro Toco site was graciously provided by
Hern\'an Quintana; Ted Griffith and Eugene Ortiz helped in field.
Ray Blundell and colleagues loaned us a C-band amp at a critical time.
Angela Qualls made figures for this paper and helped the project 
on innumerable occasions. Harvey Moseley made the
connection between time domain beam synthesis and Fourier
transform spectroscopy that we note in the paper.

Neither experiment would have been possible
without NRAO's detector development. Additionally, NRAO's
site monitoring was invaluable for assessing Cerro Toco.
We thank the NSBF for two wonderful balloon launches.  
The WOMBAT foregrounds compilation greatly assisted us in the
data analysis. We also
thank Lucent Technologies for donating the radar trailer.  This
work was supported by an NSF NYI award, a Cottrell Award from the
Research Corporation, a David and Lucile Packard Fellowship to
LP; a NASA GSRP fellowship, Dodds Fellowship, and Hubble Fellowship
to AM; a NSF graduate fellowship to MN; a Dicke Fellowship to ET;
a Sloan Foundation Award, NSF Career award (AST-9732960), to MD;  
NSF grants PHY-9222952, PHY-9600015, PHY-0099493, NASA grant
NAG5-6034, and the University of Pennsylvania. The data on which this 
paper are based are public and may be found at http://imogen.princeton.edu/mat/.

\end{document}